\documentclass[11pt,feqno]{article}
\usepackage{amssymb,bbm,amsmath,amsthm,amscd}
\usepackage{amsmath,amsthm,amsfonts,amssymb}
\usepackage{amssymb, epsfig,amssymb, latexsym, xcolor}
\usepackage{amsfonts,psfrag,color,url,amsthm}
\usepackage{vmargin}
\usepackage{graphicx}
\usepackage{subfigure}
\usepackage{algorithm}
\usepackage[font=small,labelfont=bf,tableposition=top]{caption}
\usepackage{lineno,hyperref}
\usepackage{pgfplots}
\usepackage{verbatim}
\usepackage{algpseudocode}
\usepackage{mathtools} 
\usepackage{mathrsfs} 
\usepackage[active]{srcltx} 
\usepackage{enumerate}
\ifpdf
\DeclareGraphicsExtensions{.eps,.pdf,.png,.jpg}
\else
\DeclareGraphicsExtensions{.eps}
\fi

\newtheorem{theorem}{Theorem}[section]

\newtheorem{remark}[theorem]{Remark}

\usetikzlibrary[shapes.geometric] % ConTEXt when using Tik Z
\usetikzlibrary[mindmap]

\begin{document}
\author{
B. Koc \thanks{IFP Energies Nouvelles, Rueil-Malmaison, France. {\tt birgul.koc@ifpen.fr}}, 
S. Rubino \thanks{Departamento EDAN \& IMUS, Universidad de Sevilla, Spain. {\tt samuele@us.es}},
T. Chacón Rebollo \thanks{Departamento EDAN \& IMUS, Universidad de Sevilla, Spain. {\tt chacon@us.es}},
T. Iliescu \thanks{Department of Mathematics, Virginia Tech, Blacksburg, USA. {\tt iliescu@vt.edu}}
}

\title{Residual Data-Driven Variational Multiscale Reduced Order Models for Parameter Dependent Problems}
\date{}
\maketitle

\begin{abstract}
In this paper, we investigate the modeling of sub-scale components of proper orthogonal decomposition reduced order models (POD-ROMs) of convection-dominated flows. We propose ROM closure models that depend on the ROM residual.
We illustrate the new residual-based data-driven ROM closure within the variational multiscale (VMS) framework and investigate it in the numerical simulation of a one-dimensional parameter-dependent convection-dominated convection-diffusion problem. For comparison purposes, we also investigate a  streamline-upwind Petrov-Galerkin (SUPG) ROM stabilization strategy and the standard Galerkin ROM (G-ROM).
Our numerical investigation shows that the new residual-based data-driven VMS-ROM is more accurate than both the standard G-ROM and the SUPG-ROM.
\end{abstract}


\section{Introduction} \label{sec:intro}
Over the last decade, several closure modeling strategies for reduced order models (ROMs) of turbulent flows have been developed (see the review in~\cite{ahmed2021closures}). These closure models include the effect of the discarded ROM basis functions, which can significantly increase the ROM accuracy in under-resolved turbulent flows. Probably the most active area of research in this field has been data-driven closure modeling, in which available data is used to construct accurate ROM closure operators. Most of the current data-driven ROM closures depend on the ROM coefficients.

In this paper, we investigate the modeling of sub-scale components of proper orthogonal decomposition reduced order models (POD-ROMs) of convection-dominated flows. We find our primary motivation in the low decay of the POD eigenvalues for high-Reynolds number, turbulent flows. As a consequence, the POD-ROM solution of these flows needs reduced spaces of very high dimension, incurring a high computational cost and yielding no gain with respect to full order models (FOMs), i.e., computational models obtained with classical numerical methods (e.g., finite element (FE) method).

As in the FOM case, the issue of ROM sub-scale component modeling for convection-dominated flows is closely related to the stabilization of ROMs. This is due to the diffusive effect of the sub-grid scales on the resolved scales~\cite{CSB03}. This diffusive effect has been leveraged in constructing ROM stabilized models, see, e.g.~\cite{AzaiezJCP21,ballarin2015supremizer,bergmann2009enablers,caiazzo2014numerical,ChaconCMAME22,giere2015supg,iliescu2013variational,iliescu2014variational,NovoRubinoSINUM21,pacciarini2014stabilized,reyes2020projection,stabile2018finite}. In this paper, we propose a new residual-based data-driven VMS (variational multiscale)-ROM closure framework, and we compare it with both the streamline-upwind Petrov-Galerkin ROM (SUPG-ROM) stabilization strategy and the standard Galerkin ROM (G-ROM). 

VMS models \cite{Hughes98} are increasingly used as a successful approach that seeks to simulate large scale structures in turbulent flows. In the context of FOM, a recent survey of VMS models can be found in \cite{ARCME}, where a classification into essentially residual-based and not residual-based VMS models is done.

A survey of VMS-ROM closure modeling is performed in Section IV.A.5 of \cite{ahmed2021closures}. Next, we outline several VMS-ROM closure models relevant to the new residual-based data-driven VMS-ROM closure framework. In \cite{bergmann2009enablers}, a residual-based VMS model is proposed as a ROM stabilization strategy. A two-scale VMS-ROM equipped with time-dependent orthogonal sub-grid scales is developed in \cite{reyes2020projection}. A VMS-ROM closure that includes an artificial viscosity added only to the small resolved scales of the gradient is proposed in \cite{iliescu2013variational}. Numerical tests in \cite{iliescu2013variational} show the increased numerical stability and accuracy of the VMS-ROM over the standard G-ROM and illustrate the theoretical convergence rates. In particular, a problem displaying shock-like phenomena is considered (2D traveling wave) at a moderate P\'eclet number ($\nu=10^{-4}$). In \cite{iliescu2014variational}, the VMS-ROM is extended and studied for the incompressible Navier-Stokes equations. 
Recent VMS-ROM developments can be found in, e.g.,~\cite{eroglu2017modular,parish2017unified,roop2014proper,stabile2019reduced,tello2019fluid}.

The SUPG-ROM strategy~\cite{bergmann2009enablers,giere2015supg,kragel2005streamline,pacciarini2014stabilized,zoccolan2023streamline,zoccolan2023stabilized} is one of the most popular ROM stabilization techniques. Next, we briefly mention SUPG-ROM approaches relevant to the SUPG-ROM used in our numerical investigation.
In \cite{giere2015supg}, the authors present a SUPG-ROM based on POD, which is investigated theoretically and numerically for the convection-diffusion equation. In this context, it is stressed that for convection-dominated problems whose solutions have a steep internal layer, using a stabilized discretization is necessary when using relatively coarse meshes. We note that, in practical applications, FE meshes are significantly coarser than the width of the internal layers. Relying on the SUPG-FEM \cite{brooks1982streamline}, in \cite{giere2015supg}, the authors use both offline and online stabilization procedures to deal with the numerical instabilities of the Galerkin method in both the FOM and the ROM. We note that the SUPG-ROM involves the full residual, thus being fully consistent. The study of appropriate choices of the SUPG-ROM stabilization parameter is considered in \cite{giere2015supg}. 
Two approaches are used: one based on the underlying FE discretization and the other based on the POD truncation. Thus, the question of whether the stabilization parameter should depend on the spatial resolution of the underlying FE space or on the number of POD modes used is treated in \cite{giere2015supg} by means of numerical analysis arguments. Another SUPG-ROM strategy is considered in \cite{caiazzo2014numerical}, where the authors propose and numerically investigate a SUPG-ROM stabilization method for the convection term, together with a stabilization approach for the pressure term. An error analysis of this method in \cite{caiazzo2014numerical} was performed in \cite{ChaconCMAME22}.
Recent developments in the numerical analysis of SUPG-ROMs can be found in~\cite{john2022error}.

For completeness, we remark that not fully consistent stabilization techniques (but of optimal order with respect to the FOM interpolation), such as local projection stabilization (LPS) \cite{BB01}
have also been applied in the context of POD-ROM, see, e.g., \cite{AzaiezJCP21,NovoRubinoSINUM21,RubinoSINUM20}.

In this paper, we introduce a new type of data-driven ROM closure for convection-diffusion problems, in which the closure term is a function of the ROM residual. We perform a systematic numerical study of different sub-scale modeling strategies to increase the ROM accuracy for convection-dominated problems that show the excellent accuracy properties of the new method. It advantageously compares to several other types of data-driven ROM closures. In our numerical investigation, we consider the following models:
(i) the novel residual-based data-driven VMS-ROM (R-D2-VMS-ROM); 
(ii) the standard coefficient-based data-driven VMS-ROM (C-D2-VMS-ROM)~\cite{mou2021data,xie2018data} 
 (see also~\cite{ivagnes2023pressure,san2022variational,snyder2023data}); and 
(iii) a data-driven SUPG-ROM. The first two approaches are closure models, whereas the third is a stabilization method.
We also note that all three approaches are data-driven approaches in which a least squares problem is solved in the offline stage to determine the closure operators (for the first two approaches) or the stabilization parameter (in the third approach) that are optimal with respect to the FOM data. To ensure a fair comparison, in the SUPG-ROM construction, we provide
different strategies for the computation of the stabilization parameter, apart from the data-driven approach. Our numerical investigation shows that the new R-D2-VMS-ROM yields the most accurate results.

The outline of the paper is as follows: 
In Sections~\ref{sec:fom} and \ref{sec:g_rom}, we briefly outline the FOM and G-ROM for the parameter-dependent convection-diffusion problem, respectively. Section~\ref{sec:vms_rom} describes the small-large scale decomposition that underpins the VMS-ROM framework. Section~\ref{sec:d2_vms_rom} outlines two types of D2-VMS-ROM. Specifically, Section~\ref{sec:r_d2_vms_rom} presents the new R-D2-VMS-ROM with two different strategies, and
Section~\ref{sec:c_d2_vms_rom} describes the standard C-D2-VMS-ROM.
In Section~\ref{sec:supg_rom}, we outline the SUPG-ROM and list different options for the stabilization parameter. In Section~\ref{sec:numerical_results}, we provide a numerical investigation of four models: G-ROM, R-D2-VMS-ROM, C-D2-VMS-ROM, and SUPG-ROM. Section~\ref{sec:conclusions_outlook} concludes the paper by presenting a short summary and future research directions.

%%%%%%%%%%
\section{Full Order Order Model (FOM)}  \label{sec:fom}
As a mathematical model, we use a one-dimensional parameter-dependent convection-diffusion (CD) problem with a small diffusion coefficient:
\begin{eqnarray} \label{eqn:parameter_cd_problem}
\begin{cases}
- \mu \partial_{xx} u + c \partial_x u = f \quad \text{for} \, \, x \in [0,1], \\
u(0)= 0 \quad \text{and} \quad u(1)=0,
\end{cases}
\end{eqnarray}
where $f$ is the forcing term, $c$ the convection velocity field, $u$ the variable of interest, and $\mu$ the diffusion coefficient.

\begin{figure}[h!] 
\begin{center}
\includegraphics[width=0.32\textwidth,height=0.3\textwidth]{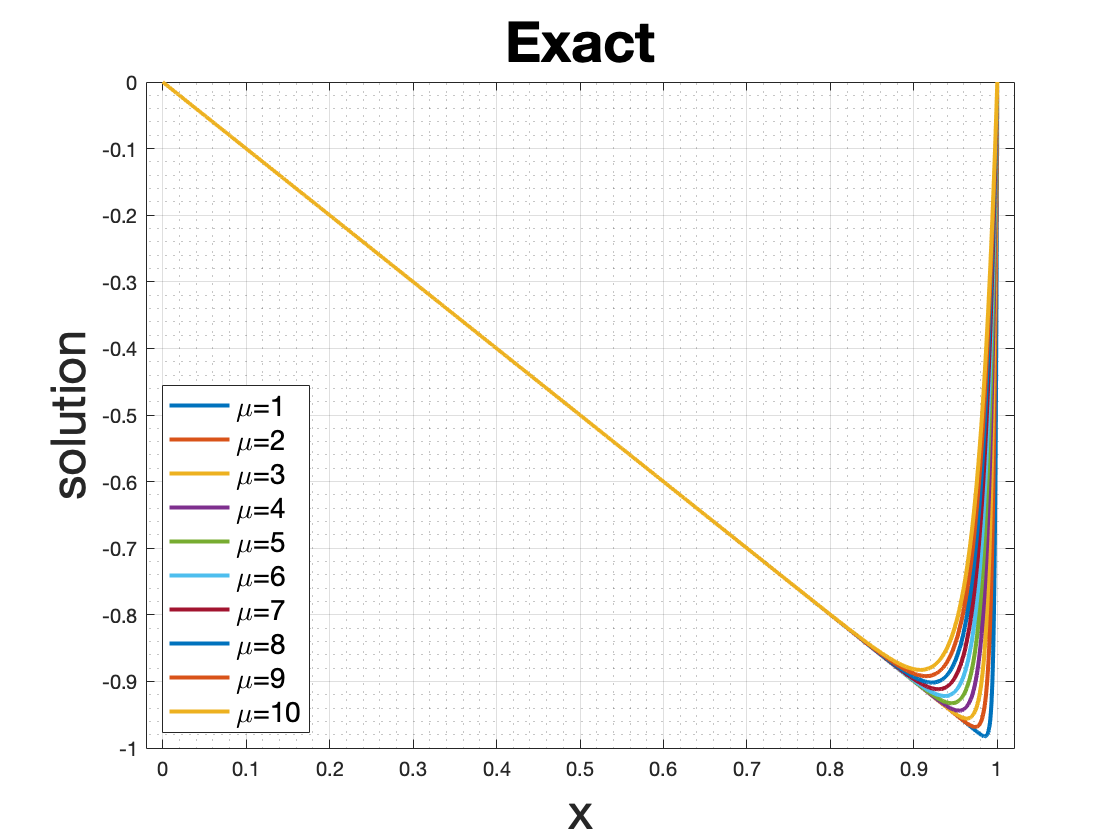}
\includegraphics[width=0.32\textwidth,height=0.3\textwidth]{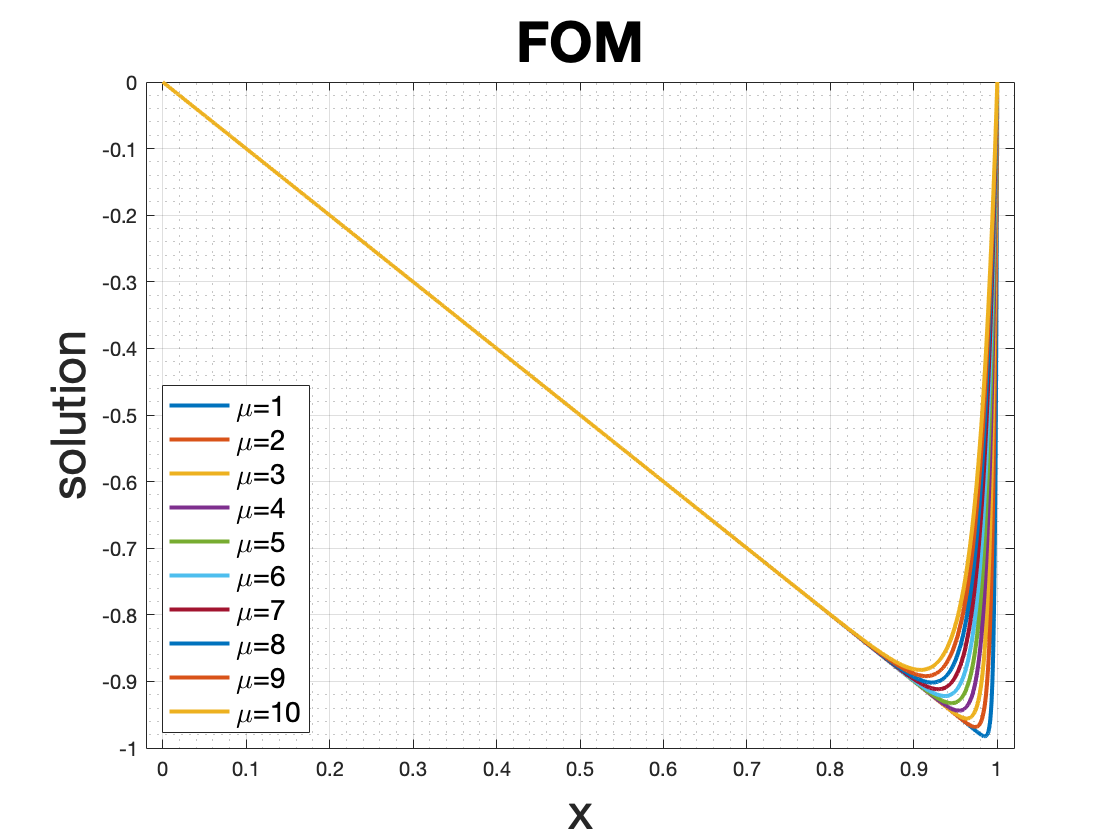}
\includegraphics[width=0.32\textwidth,height=0.3\textwidth]{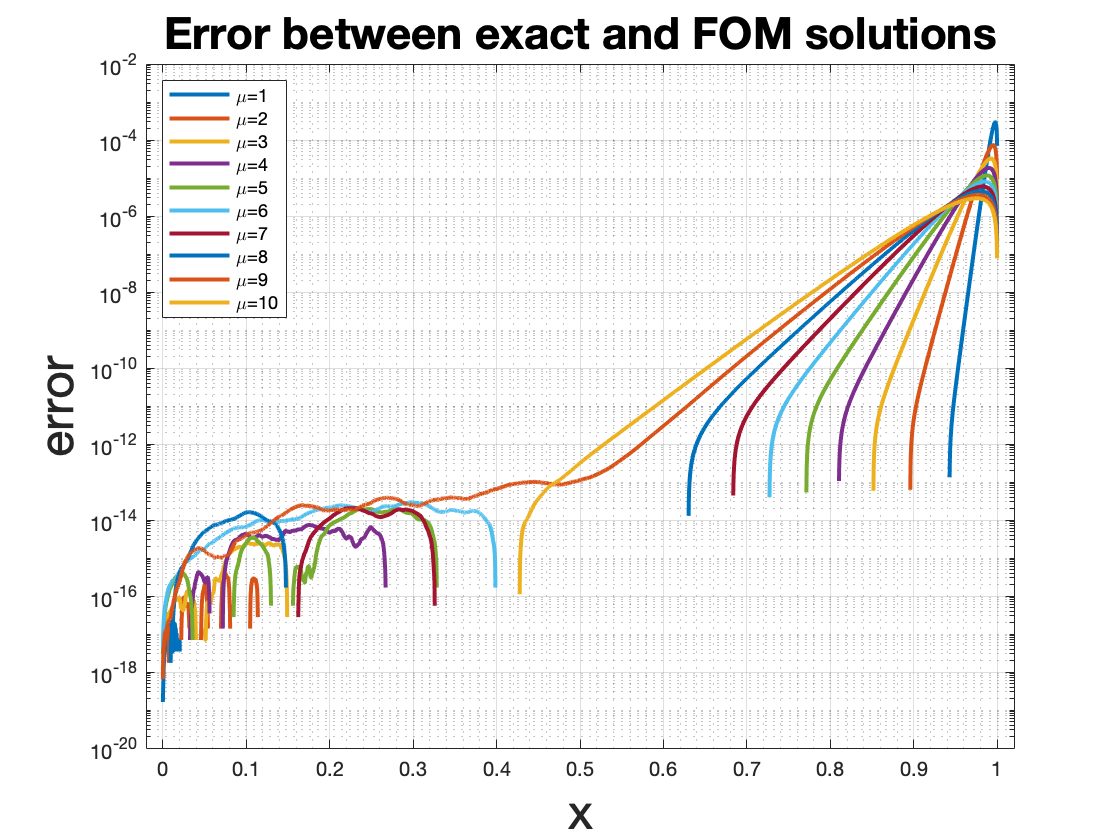}
\caption{Exact solution \eqref{eqn:exact_soln}, FOM solution of \eqref{eqn:parameter_cd_problem}, and error plots with $f=-c=-400$ and different values of the parameter $\mu$.
} \label{fig:exact_fom_plots}
\end{center} 
\end{figure}

In the parameter-dependent CD problem~\ref{eqn:parameter_cd_problem}, we vary the parameter $\mu$ to obtain different solutions. In Figure~\ref{fig:exact_fom_plots}, we plot the exact and FOM solutions, which solve \eqref{eqn:parameter_cd_problem} by using the finite element method, and the error between the exact and FOM solutions for different $\mu$ values. The error plot shows that the FOM solutions are well resolved. Thus, we will use them as snapshots in the next sections.

%%%%%%%%%%
\section{Galerkin Reduced Order Model (G-ROM)}  \label{sec:g_rom}
This section provides a brief overview of the standard G-ROM strategy. The G-ROM is one of the most common types of ROMs for fluid flows. In Algorithm~\ref{alg:g-rom}, we outline the G-ROM construction.
\begin{algorithm} 
\caption{Galerkin ROM (G-ROM)}
\label{alg:g-rom}
\begin{algorithmic}[1] 
\State Use available FOM data to construct dominant modes by using the POD, $\{ \boldsymbol \varphi_{1}, \ldots, \boldsymbol \varphi_{L}\}$, $L \ll d$ (where $d$ is the dimension of the input dataset), which correspond to the largest relative kinetic energy content and represent the dominant spatial structures of the given test problem;

\vspace{3mm}

\State Construct a ROM approximation:
\begin{align} \label{eqn:rom_soln_L_dim}
\boldsymbol u_L = \sum_{j=1}^{L} (\boldsymbol a_L)_j \, \boldsymbol \varphi_j,
\end{align}
as a linear combination of ROM basis functions $\boldsymbol \varphi_j$ with ROM coefficients $(\boldsymbol a_L)_j$;

\vspace{3mm}

\State Replace $\boldsymbol u$ in the given test problem~\eqref{eqn:parameter_cd_problem} with the ROM solution $\boldsymbol u_L$ given in \eqref{eqn:rom_soln_L_dim}; 

\vspace{3mm}

\State Use the Galerkin projection, which projects the system obtained in step 3 onto the ROM space $ \boldsymbol X^L$ spanned by $\{ \boldsymbol \varphi_{1}, \ldots, \boldsymbol \varphi_{L}\}$.
\end{algorithmic}
 \label{alg:g_rom}
\end{algorithm}

By using Algorithm~\ref{alg:g_rom} for the CD problem~\eqref{eqn:parameter_cd_problem}, we obtain the following G-ROM: 
\begin{align} \label{eqn:g_rom}
\boldsymbol  A_{LL} \, \boldsymbol a_L = \boldsymbol b_L,
\end{align}
where, for $i,j=1,\ldots,L$, the ROM operators are the matrix $(\boldsymbol A_{LL})_{ij}:=  \mu \langle \partial_x \boldsymbol \varphi_j, \partial_x \boldsymbol \varphi_i\rangle + \langle c \, \partial_x \boldsymbol \varphi_j, \boldsymbol \varphi_i\rangle$, and the vector $( \boldsymbol b_L)_i:= \langle\boldsymbol f, \boldsymbol \varphi_i\rangle$, where $\langle \cdot, \cdot \rangle$ represents the $L^2$ inner product, and $\boldsymbol a_L$ is the vector of G-ROM coefficients $(\boldsymbol a_L)_j$, which needs to be determined.

%%%%%%%%
\section{Variational Multiscale Reduced Order Model (VMS-ROM)} \label{sec:vms_rom}
In this section, we construct the VMS-ROM framework, which will be used in the next sections to build data-driven VMS-ROMs.

First, we note that when all the ROM modes are used to create a ROM solution, the ROM approximation reads
\begin{align} \label{eqn:rom_soln_d_dim}
\boldsymbol u_d = \sum_{j=1}^{d} (\boldsymbol a_d)_j \, \boldsymbol \varphi_j.
\end{align}
In this case, $\boldsymbol u_d$ is the most accurate ROM approximation of the FOM solution with the given data in the POD sense (i.e., from the energetic point of view \cite{volkwein2013proper}).

For laminar flows, using a few ($L \ll d $) ROM basis functions is enough to capture the main dynamics of the given problem ({\it resolved regime}). In other words, a low-dimensional ROM solution $\boldsymbol u_L$, with small $L \ll d $, yields an accurate approximation of the FOM solution. 

However, for turbulent flows, the low-dimensional ROM solution \eqref{eqn:rom_soln_L_dim} with small $L \ll d $ is not an accurate approximation of the FOM solution ({\it under-resolved regime}). To increase the accuracy of the $L$-dimensional ROM solution~\eqref{eqn:rom_soln_L_dim}, we roughly have two options: 
(i) increase the G-ROM dimension, $L$, or 
(ii) add a low-dimensional closure term to the G-ROM. In this paper, we aim to increase the numerical accuracy without significantly increasing the computational cost. Thus, we choose the second option. Next, we explain what ROM closure modeling is (see~\cite{ahmed2021closures} for a review) and how it is performed in a VMS setting.

The orthogonality of the ROM basis functions (since we used the POD to construct the basis) allows us to decompose the ROM space as follows: $\boldsymbol X^d = \boldsymbol X^L \oplus \boldsymbol X^S$, where $\boldsymbol X^d:= \text{span}\{\boldsymbol \varphi_1,..., \boldsymbol \varphi_d \}$, $\boldsymbol X^L:= \text{span}\{\boldsymbol \varphi_1,...,\boldsymbol \varphi_L\}$, and $\boldsymbol X^S:= \text{span}\{\boldsymbol \varphi_{L+1},..., \boldsymbol \varphi_d\}$. By using the same decomposition, we define the large and sub-scale solutions of the most accurate (in the previously mentioned sense) ROM solution, $\boldsymbol u_d$, as follows:
\begin{subequations}
\begin{align}
\boldsymbol u_L  &:= \sum_{j=1}^{L} (\boldsymbol a_L)_j \, \boldsymbol \varphi_j, \\
\boldsymbol u_S & := \sum_{j=L+1}^{d} (\boldsymbol a_S)_j \, \boldsymbol \varphi_j.
\end{align}
\end{subequations}

The most accurate ROM approximation in \eqref{eqn:rom_soln_d_dim}, $\boldsymbol u_d$, solves the following bilinear-linear form of the CD problem \eqref{eqn:parameter_cd_problem}  
\begin{align} \label{eqn:general_form_d}
a(\boldsymbol u_d, \boldsymbol v_d ) = \langle \boldsymbol f, \boldsymbol v_d \rangle, \quad \forall \,  \boldsymbol v_d \in \boldsymbol X^d,
\end{align}
where the bilinear form $a(\boldsymbol u_d, \boldsymbol v_d):= -\mu \langle \partial_{xx} \boldsymbol u_d, \boldsymbol v_d \rangle + \langle c \partial_x \boldsymbol u_d, \boldsymbol v_d \rangle$. By using the VMS method and choosing $\boldsymbol v_\ell = \boldsymbol \varphi_\ell \in \boldsymbol X^L$ and $\boldsymbol v_s = \boldsymbol \varphi_s \in \boldsymbol X^S$ $(\boldsymbol v_d = \boldsymbol v_\ell + \boldsymbol v_s )$, we can decompose \eqref{eqn:general_form_d} into two problems as follows:
\begin{subequations}
\begin{align}
a(\boldsymbol u_L, \boldsymbol \varphi_\ell ) &= \langle \boldsymbol f, \boldsymbol \varphi_\ell \rangle - a(\boldsymbol u_S, \boldsymbol \varphi_\ell )= \langle Res(\boldsymbol u_S), \boldsymbol \varphi_\ell \rangle := Res_{\boldsymbol L}(\boldsymbol u_S), \label{eqn:general_form_large} \\
a(\boldsymbol u_S, \boldsymbol \varphi_s ) &= \langle \boldsymbol f,  \boldsymbol \varphi_s \rangle - a(\boldsymbol u_L, \boldsymbol \varphi_s )= \langle Res(\boldsymbol u_L), \boldsymbol \varphi_s  \rangle := Res_{\boldsymbol S}(\boldsymbol u_L), \label{eqn:general_form_small}
\end{align}
\end{subequations}
where $Res(\boldsymbol u) = \boldsymbol f - ( -\mu \partial_{xx} \boldsymbol u + c \partial_x \boldsymbol u )$. The matrix-vector forms of \eqref{eqn:general_form_large}-\eqref{eqn:general_form_small} are as follows:
\begin{subequations}
\begin{align}
\boldsymbol A_{LL} \, \boldsymbol a_L + \boldsymbol A_{LS} \, \boldsymbol a_S = \boldsymbol b_L,  \label{eqn:matrix_vector_form_large} \\
\boldsymbol A_{SL} \, \boldsymbol a_L + \boldsymbol A_{SS} \, \boldsymbol a_S = \boldsymbol b_S. \label{eqn:matrix_vector_form_small}
\end{align}
\end{subequations}

The matrices and vectors in \eqref{eqn:matrix_vector_form_large}--\eqref{eqn:matrix_vector_form_small} are defined as follows:
\begin{align}\label{eqn:matrix_vector_def}
\begin{aligned}
\boldsymbol (\boldsymbol A_{MN})_{ij} & = \mu \, (\boldsymbol S_{MN})_{ij} + (\boldsymbol C_{MN})_{ij}, \\
(\boldsymbol b_M)_i & = \langle \boldsymbol f, \boldsymbol \varphi_i \rangle, \\
(\boldsymbol S_{MN})_{ij} & = \langle \partial_x \boldsymbol \varphi_j, \partial_x \boldsymbol \varphi_i \rangle, \\
(\boldsymbol C_{MN})_{ij} & = \langle c\,\partial_x \boldsymbol \varphi_j, \boldsymbol \varphi_i \rangle,
\end{aligned}
\end{align}
where $M,N$ could be either $L$ (and the corresponding indices vary from $1$ to $L$) or $S$ (and the corresponding indices vary from $L+1$ to $d$).

The VMS-ROM idea can be explained by using the matrix formulation in~\eqref{eqn:matrix_vector_form_large}--\eqref{eqn:matrix_vector_form_small}.
First, we note that this matrix form was obtained by using the variational formulation in~\eqref{eqn:general_form_large}--\eqref{eqn:general_form_small}. 
Second, we note that the matrix formulation in~\eqref{eqn:matrix_vector_form_large}--\eqref{eqn:matrix_vector_form_small} is a multiscale formulation since $\boldsymbol a_L$ and $\boldsymbol a_S$ correspond to the large and small scales in the system, respectively.
Thus, the matrix formulation in~\eqref{eqn:matrix_vector_form_large}--\eqref{eqn:matrix_vector_form_small} is truly a variational multiscale ROM formulation.

\begin{figure} 
\begin{center}
\begin{tikzpicture}
\draw [|<->|] (-0.6,-0.5) -- node[left=0.1mm] {$L$} (-0.6,0.5);
\draw (0,0) node[minimum height=1cm,minimum width=1cm,draw,fill=blue!80] {$\boldsymbol A_{LL}$};
\draw [|<->|] (-0.5,0.6) -- node[above=0.1mm] {$L$} (0.5,0.6);
\draw (1.5,0) node[minimum height=1cm,minimum width=2cm,draw,fill=blue!50] {$\boldsymbol A_{LS}$};
\draw (0,-1.5) node[minimum height=2cm,minimum width=1cm,draw,fill=blue!50] {$\boldsymbol A_{SL}$};
\draw [|<->|] (0.5,-2.65) -- node[below=0.1mm] {$S$} (2.5,-2.65);
\draw (1.5,-1.5) node[minimum height=2cm,minimum width=2cm,draw,fill=red!80] {$\boldsymbol A_{SS}$};
\draw [|<->|] (2.65,-2.5) -- node[right=0.1mm] {$S$} (2.65,-0.5);
\draw (3.65,0) node[minimum height=1cm,minimum width=0.25cm,draw,fill=blue!80] {$\boldsymbol a_L$};
\draw [|<->|] (4.1,-0.5) -- node[right=1mm] {$L$} (4.1,0.5);
\draw (3.65,-1.5) node[minimum height=2cm,minimum width=0.25cm,draw,fill=red!80] {$\boldsymbol a_S$};
\draw [|<->|] (4.1,-2.5) -- node[right=1mm] {$S$} (4.1,-0.5);
\draw [] (5.4,-0.5)  node[left=1mm] {$=$} (5.45,-0.5);
\draw (6,0) node[minimum height=1cm,minimum width=0.5cm,draw,fill=blue!80] {$\boldsymbol b_L$};
\draw [|<->|] (6.425,-0.5) -- node[right=1mm] {$L$} (6.425,0.5);
\draw (6,-1.5) node[minimum height=2cm,minimum width=0.5cm,draw,fill=red!80] {$\boldsymbol b_S$};
\draw [|<->|] (6.425,-2.5) -- node[right=1mm] {$S$} (6.425,-0.5);
\end{tikzpicture}
\caption{Matrix-vector diagram of the coupled system \eqref{eqn:matrix_vector_form_large}-\eqref{eqn:matrix_vector_form_small}.} \label{fig:coupled_system_diagram}
\end{center} 
\end{figure}
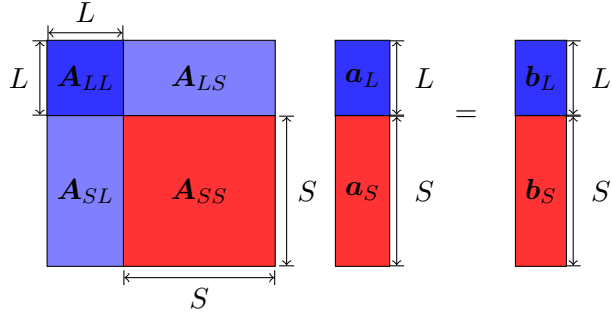

The main goal in VMS-ROM is to find an accurate approximation of $\boldsymbol a_L$ \textit{without solving for $\boldsymbol a_S$}.
The rationale is that, since $L \ll d$,  $\boldsymbol a_L$ can be computed efficiently.
In contrast, since $L \ll S$, we should try to avoid the expensive computation of $\boldsymbol a_S$.
The challenge, however, is that the equations for $\boldsymbol a_L$ and $\boldsymbol a_S$ in~\eqref{eqn:matrix_vector_form_large}--\eqref{eqn:matrix_vector_form_small} are coupled.
Of course, one could decouple the two equations by using the Schur complement, but that would involve the computation of $\boldsymbol A_{SS}^{-1}$, which is expensive (since $S$ is large), see Figure~\ref{fig:coupled_system_diagram}.

The VMS-ROM strategy centers around a simple idea:
VMS-ROMs reduce the large system of equations~\eqref{eqn:matrix_vector_form_large}--\eqref{eqn:matrix_vector_form_small} to a low-dimensional equation for $\boldsymbol a_L$. 
Figure~\ref{fig:coupled_system_diagram} represents a schematic representation of the VMS-ROM idea.
To obtain an accurate approximation for $\boldsymbol a_L$, the effect of $\boldsymbol a_S$ needs to be modeled.
In the next section, we present two new data-driven strategies for modeling the effect of $\boldsymbol a_S$.

\section{Data-Driven Variational Multiscale ROM (D2-VMS-ROM)} \label{sec:d2_vms_rom}
In this section, we present two fundamentally different types of D2-VMS-ROMs.
In Section~\ref{sec:r_d2_vms_rom}, we propose two novel D2-VMS-ROMs in which the effect of sub-scales on the large resolved scales is modeled by using the large scale residual.
These two new models are denoted by R1-D2-VMS-ROM and R2-D2-VMS-ROM.
For comparison purposes, in Section~\ref{sec:c_d2_vms_rom}, we outline the construction of the recently proposed D2-VMS-ROM, in which the effect of sub-scales on the large resolved scales is modeled by using the large scale ROM coefficients, $\boldsymbol a_L$. 
This model is denoted by C-D2-VMS-ROM.

%%%%%%%%%%%%%%%%%%%%%%%%%%%%%%%%
\subsection{Residual-Based Data-Driven Variational Multiscale ROM (R-D2-VMS-ROM) } \label{sec:r_d2_vms_rom}
In this section, we propose two different R-D2-VMS-ROM strategies, i.e., R1-D2-VMS-ROM and R2-D2-VMS-ROM. 

To construct R1-D2-VMS-ROM, we use only one ansatz to model the sub-scales in \eqref{eqn:matrix_vector_form_large}, $\boldsymbol a_S$, as a function of the residual term $Res_{\boldsymbol S}(\boldsymbol a_L):= \big( \boldsymbol b_S - \boldsymbol A_{SL} \, \boldsymbol a_L \big)$.
The idea behind this strategy is that we avoid solving the expensive, high-dimensional equation~\eqref{eqn:matrix_vector_form_small}. 
Instead, we only leverage the information in~\eqref{eqn:matrix_vector_form_small} (i.e., the fact that $\boldsymbol a_S$ depends on the residual $Res_{\boldsymbol S}(\boldsymbol a_L)$) to model the sub-scales in \eqref{eqn:matrix_vector_form_large}. 

To construct R2-D2-VMS-ROM, we use two ansatzes.
The first ansatz is the same as the ansatz used to construct R1-D2-VMS-ROM:
We model the sub-scales in \eqref{eqn:matrix_vector_form_large}, $\boldsymbol a_S$, as a function of the residual term $Res_{\boldsymbol S}(\boldsymbol a_L):= \big( \boldsymbol b_S - \boldsymbol A_{SL} \, \boldsymbol a_L \big)$.
The second ansatz models the residual $Res_{\boldsymbol L}( \boldsymbol a_S):= \big( \boldsymbol b_L - \boldsymbol A_{LS} \, \boldsymbol a_S \big)$ by using $Res_{\boldsymbol L}( \boldsymbol a_S^{approx})$, where $\boldsymbol a_S^{approx}$ is the sub-scale approximation in the first ansatz.
We emphasize that, just as we did in the R1-D2-VMS-ROM construction, we are not solving the expensive, high-dimensional equation~\eqref{eqn:matrix_vector_form_small}.
Instead, we only leverage the information in~\eqref{eqn:matrix_vector_form_small} to postulate our ansatzes.

Depending on how we model the sub-scales in R1-D2-VMS-ROM and R2-D2-VMS-ROM (i.e., if we use a linear or an affine ansatz), we present four different R-D2-VMS-ROMs. Since both R1-D2-VMS-ROM and R2-D2-VMS-ROM contain more information from the sub-scales in \eqref{eqn:matrix_vector_form_large} and \eqref{eqn:matrix_vector_form_small} and are constructed by using the residual ansatz to model the closure terms, the expectation is that R-D2-VMS-ROMs are both more accurate and more consistent than standard D2-VMS-ROMs, i.e., C-D2-VMS-ROMs (which are presented in Section~\ref{sec:c_d2_vms_rom}).

%%%%%%%%%%%%%%%%%%%%%%%%%%%%%%%%%
\subsubsection{The new R1-D2-VMS-ROM strategy}  \label{sec:r1_d2_vms_rom}
The following two R1-D2-VMS-ROM variants are obtained by using only one ansatz to model the sub-scales $\boldsymbol a_S$ in \eqref{eqn:matrix_vector_form_large}:
\begin{subequations}
\begin{align}
\boldsymbol a_S \, \approx \, \textbf{Ansatz}(Res_{\boldsymbol S}(\boldsymbol a_L)) = \widetilde{\boldsymbol A} \, Res_{\boldsymbol S}(\boldsymbol a_L), \label{eqn:residual_based_ansatz1a} \\
\textbf{R1a-ROM}: \, \big( \boldsymbol A_{LL} - \boldsymbol A_{LS} \, \widetilde{\boldsymbol A} \, \boldsymbol A_{SL} \big) \, \boldsymbol a_L = \boldsymbol b_L - \boldsymbol A_{LS} \, \widetilde{\boldsymbol A} \, \boldsymbol b_S.  \label{eqn:r1a_d2_vms_rom}
\end{align}
\end{subequations}

\begin{subequations}
\begin{align}
\boldsymbol a_S \, \approx \, \textbf{Ansatz}(Res_{\boldsymbol S}(\boldsymbol a_L)) = \widetilde{\boldsymbol A} \, Res_{\boldsymbol S}(\boldsymbol a_L) + \widetilde{\boldsymbol b}, \label{eqn:residual_based_ansatz1b} \\
\textbf{R1b-ROM}: \, \big( \boldsymbol A_{LL} - \boldsymbol A_{LS} \, \widetilde{\boldsymbol A} \, \boldsymbol A_{SL} \big) \, \boldsymbol a_L = \boldsymbol b_L - \boldsymbol A_{LS} \, \widetilde{\boldsymbol A} \, \boldsymbol b_S - \boldsymbol A_{LS} \, \widetilde{\boldsymbol b}. \label{eqn:r1b_d2_vms_rom}
\end{align}
\end{subequations}

The main difference between the ansatzes in \eqref{eqn:residual_based_ansatz1a} and \eqref{eqn:residual_based_ansatz1b} is the D2 operator $\widetilde{\boldsymbol b}$. 
To ensure a fair comparison of these two methods, in Section~\ref{sec:numerical_results}, we provide the consistency error tables to determine whether the vector 
$\widetilde{\boldsymbol b}$ is needed or not.

For both cases, to construct the D2 operators in \eqref{eqn:r1a_d2_vms_rom} and \eqref{eqn:r1b_d2_vms_rom}, we need to solve 
the following least squares problem:
\begin{align} \label{eqn:d2_method_residual1}
\min_{\text{D2 operators}}   \quad \sum_{k=1}^{M} \left \|  \, \boldsymbol a_S^k - \textbf{Ansatz}\big( Res_{\boldsymbol S}(\boldsymbol a_L^k) \big) \, \right \|^2_{L^2},
\end{align} 
where $M$ represents the number of snapshots. 
After solving \eqref{eqn:d2_method_residual1} for the D2 operators $\widetilde{\boldsymbol A}$ and $\widetilde{\boldsymbol b}$ for the ansatzes \eqref{eqn:residual_based_ansatz1a} and \eqref{eqn:residual_based_ansatz1b}, and plugging them into \eqref{eqn:matrix_vector_form_large}, we get the models R1a-ROM \eqref{eqn:r1a_d2_vms_rom} and R1b-ROM \eqref{eqn:r1b_d2_vms_rom}.\\

\subsubsection{The new R2-D2-VMS-ROM strategy}  \label{sec:r2_d2_vms_rom}

Next, we present two more R-D2-VMS-ROM variants, which are derived by using two ansatzes and are denoted by R2-D2-VMS-ROMs. The main differences between the R1 and R2 types of D2-VMS-ROMs are the following: (i) the R2 models use two ansatzes whereas the R1 models use only one ansatz, and (ii) because the R2 models use more ansatzes, they have more information related to the sub-scales since they gradually model the sub-scales.

In R2-D2-VMS-ROMs, the first ansatz is constructed to approximate the sub-scales $\boldsymbol a_S$ in \eqref{eqn:matrix_vector_form_large} by using the residual $Res_{\boldsymbol S}(\boldsymbol a_L)$ in \eqref{eqn:matrix_vector_form_small}, as done in \eqref{eqn:residual_based_ansatz1a} and \eqref{eqn:residual_based_ansatz1b} for R1a-ROM and R1b-ROM, respectively. We solve the least square problem \eqref{eqn:d2_method_residual1} to obtain the D2 operators $\widetilde{\boldsymbol A_1}$ and $\widetilde{\boldsymbol b_1}$ for the ansatzes \eqref{eqn:residual_based_ansatz2_1a} and \eqref{eqn:residual_based_ansatz2_2a}.

The second ansatz in R2-D2-VMS-ROM is used to approximate the residual $Res_{\boldsymbol L}(\boldsymbol a_S):=\big( \boldsymbol b_L - \boldsymbol A_{LS} \, \boldsymbol a_S \big)$ in \eqref{eqn:matrix_vector_form_large} by using the approximated residual sub-scales $Res_{\boldsymbol L}(\boldsymbol a_S^{approx})$, which is obtained by using the first ansatzes \eqref{eqn:residual_based_ansatz2_1a} and \eqref{eqn:residual_based_ansatz2_2a}. After solving the least square problem \eqref{eqn:dd_method2b}, we obtain the D2 operators $\widetilde{\boldsymbol A_2}$ and $\widetilde{\boldsymbol b_2}$.
\begin{align} \label{eqn:dd_method2b}
\begin{aligned}
\min_{\text{D2 operators}}   \quad \sum_{k=1}^{M} \left \| \, Res_{\boldsymbol L}(\boldsymbol a_S^k) - \textbf{Ansatz2}(Res_{\boldsymbol L}(\boldsymbol a_S^{approx,k})) \, \right \|^2_{L^2}.
\end{aligned}
\end{align}

Using the operators $\widetilde{\boldsymbol A_1}$ and $\widetilde{\boldsymbol b_1}$, and $\widetilde{\boldsymbol A_2}$ and $\widetilde{\boldsymbol b_2}$, yields the following R2-D2-VMS-ROMs, i.e., R2a-ROM \eqref{eqn:r2_1a_d2_vms_rom} and R2b-ROM \eqref{eqn:r2_2a_d2_vms_rom}:
\begin{subequations}
\begin{align}
\boldsymbol a_S \, \approx \, \boldsymbol a_S^{approx} = \textbf{Ansatz1}(Res_{\boldsymbol S}(\boldsymbol a_L)) = \widetilde{\boldsymbol A}_1 \, Res_{\boldsymbol S}(\boldsymbol a_L), \label{eqn:residual_based_ansatz2_1a} \\
Res_{\boldsymbol L}( \boldsymbol a_S) \, \approx \, \textbf{Ansatz2}(Res_{\boldsymbol L}(\boldsymbol a_S^{approx})) =  \widetilde{\boldsymbol A}_2 \, Res_{\boldsymbol L}( \boldsymbol a_S^{approx} ), \label{eqn:residual_based_ansatz2_1b} \\
\textbf{R2a-ROM}: \, \big( \boldsymbol A_{LL} - \widetilde{\boldsymbol A}_2  \, \boldsymbol A_{LS} \, \widetilde{\boldsymbol A}_1 \, \boldsymbol A_{SL} \big) \, \boldsymbol a_L = \widetilde{\boldsymbol A}_2 \, (\boldsymbol b_L - \boldsymbol A_{LS} \, \widetilde{\boldsymbol A}_1 \, \boldsymbol b_S ).  \label{eqn:r2_1a_d2_vms_rom}
\end{align}
\end{subequations}

\begin{subequations}
\begin{align}
\boldsymbol a_S \, \approx \, \textbf{Ansatz1}(Res_{\boldsymbol S}(\boldsymbol a_L)) =  \widetilde{\boldsymbol A}_1 \, Res_{\boldsymbol S}(\boldsymbol a_L) + \widetilde{\boldsymbol b}_1, \label{eqn:residual_based_ansatz2_2a} \\
Res_{\boldsymbol L}( \boldsymbol a_S) \, \approx \, \textbf{Ansatz2}(Res_{\boldsymbol L}(\boldsymbol a_S^{approx})) = \widetilde{\boldsymbol A}_2 \, Res_{\boldsymbol L}( \boldsymbol a_S^{approx} ) + \widetilde{\boldsymbol b}_2, \label{eqn:residual_based_ansatz2_2b} \\
\textbf{R2b-ROM}: \, \big( \boldsymbol A_{LL} - \widetilde{\boldsymbol A}_2  \, \boldsymbol A_{LS} \, \widetilde{\boldsymbol A}_1 \, \boldsymbol A_{SL} \big) \, \boldsymbol a_L = \widetilde{\boldsymbol A}_2 \, \Big[ \boldsymbol b_L - \boldsymbol A_{LS} \, ( \widetilde{\boldsymbol A}_1 \, \boldsymbol b_S + \widetilde{\boldsymbol b}_1 ) \Big] + \widetilde{\boldsymbol b}_2.  \label{eqn:r2_2a_d2_vms_rom}
\end{align}
\end{subequations}

\subsection{Coefficient-Based Data-Driven Variational Multiscale ROM (C-D2-VMS-ROM)} \label{sec:c_d2_vms_rom} 
This section recalls the standard C-D2-VMS-ROMs \cite{mou2021data}. In this method, we use a linear ansatz to model the sub-scale contribution in \eqref{eqn:matrix_vector_form_large}, $\boldsymbol A_{LS} \, \boldsymbol a_S$, as a function of the large-scale ROM coefficients, $\boldsymbol a_L$, i.e., $\textbf{Ansatz}(\boldsymbol a_L)$. Then, we consider two different ansatzes, which yield two different C-D2-VMS-ROMs:
\begin{subequations}
\begin{align} 
& \boldsymbol A_{LS} \, \boldsymbol a_S \, \approx \, \textbf{Ansatz}(\boldsymbol a_L) =  \widetilde{\boldsymbol A} \,  \boldsymbol a_L, \label{eqn:coeff_based_ansatz1a} \\  
\textbf{C1a-ROM}: \quad & \big( \boldsymbol A_{LL} +  \widetilde{\boldsymbol A} \big) \, \boldsymbol a_L = \boldsymbol b_L. \label{eqn:c1a_d2_vms_rom}
\end{align}
\end{subequations} 

\begin{subequations}
\begin{align} 
& \boldsymbol A_{LS} \, \boldsymbol a_S \, \approx \, \textbf{Ansatz}(\boldsymbol a_L) = \widetilde{\boldsymbol A} \,  \boldsymbol a_L  + \widetilde{ \boldsymbol b}, \label{eqn:coeff_based_ansatz1b} \\  
\textbf{C1b-ROM}: \quad & \big( \boldsymbol A_{LL} +  \widetilde{\boldsymbol A} \big) \, \boldsymbol a_L = \boldsymbol b_L -  \widetilde{\boldsymbol b}. \label{eqn:c1b_d2_vms_rom}
\end{align}
\end{subequations} 

The expectation is that the C1b-ROM yields more accurate results than the C1a-ROM since the ansatz in the C1b-ROM has more flexibility, i.e., it contains the D2 operator $\widetilde{\boldsymbol b}$, whereas the C1a-ROM does not.

For both cases, to construct the D2 operators in \eqref{eqn:c1a_d2_vms_rom} and \eqref{eqn:c1b_d2_vms_rom}, we need to solve the following least squares problem:
\begin{align} \label{eqn:d2_method_coefficient}
\min_{\text{D2 operators}}  \quad \sum_{k=1}^{M} \left \| \, \boldsymbol A_{LS} \, \boldsymbol a_S^k - \textbf{Ansatz}(\boldsymbol a_L^k) \, \right \|^2_{L^2}.
\end{align} 
After solving \eqref{eqn:d2_method_coefficient} for the D2 operators
$\widetilde{\boldsymbol A}$ and $\widetilde{\boldsymbol b}$ for the ansatzes \eqref{eqn:coeff_based_ansatz1a} and \eqref{eqn:coeff_based_ansatz1b}, and plugging them into \eqref{eqn:matrix_vector_form_large}, 
we get the C1a-ROM as \eqref{eqn:c1a_d2_vms_rom} and C1b-ROM as \eqref{eqn:c1b_d2_vms_rom}.

%%%%%%%%%%%%%
\section{Streamline-Upwind Petrov–Galerkin ROM (SUPG-ROM)} \label{sec:supg_rom}
This section presents the SUPG-ROM for \eqref{eqn:parameter_cd_problem}. The SUPG-ROM is obtained by adding a stabilization term to the G-ROM~\eqref{eqn:g_rom}, as follows:
\begin{align} \label{eqn:general_supg_rom}
a(\boldsymbol u_L, \boldsymbol \varphi_i ) + \tau_L \, \langle \, Res(\boldsymbol u_L),  c \partial_x \boldsymbol \varphi_i \, \rangle = \langle \boldsymbol f, \boldsymbol \varphi_i \rangle, \quad i=1,\ldots,L,
\end{align}
where $\tau_L$ is the so-called ``stabilization coefficient'' and $Res(\boldsymbol u_L) = \boldsymbol f -  ( -\mu \partial_{xx} \boldsymbol u_L + c \partial_x \boldsymbol u_L )$. 
From \eqref{eqn:general_supg_rom}, we obtain the following matrix-vector form for the SUPG-ROM:
\begin{align} \label{eqn:matrix_vector_supg_rom}
\textbf{SUPG-ROM($\tau_L$)}: \, ( \boldsymbol A_{LL}  + \tau_L \, \boldsymbol A_{supg} ) \, \boldsymbol a_L = \boldsymbol b_L,
\end{align}
where $\boldsymbol A_{LL} $ and $\boldsymbol b_L$ are G-ROM operators that were defined right after \eqref{eqn:g_rom}, $\tau_L \boldsymbol A_{supg}$ is the stabilization term, and $\boldsymbol a_L$ is the SUPG-ROM coefficient that has to be determined. For the derivation of $\boldsymbol A_{supg}$, for simplicity, we will use a linear FE basis and POD modes with zero boundary conditions, so that only the weak form of the convective part does not vanish in the SUPG derivation. In other words, $\boldsymbol A_{supg} =  -c^2 \, \boldsymbol S_{LL} $, where the stiffness matrix $\boldsymbol S_{LL}$ was defined in \eqref{eqn:matrix_vector_def}, and, for simplicity, we consider a constant convection velocity $c$.

We use three different strategies to find the optimal stabilization coefficient $\tau_L$ value in SUPG-ROM~\eqref{eqn:matrix_vector_supg_rom}. The first approach uses the following two formulas inspired by the FE case in \cite{christie1976finite}:
\begin{subequations}
\begin{align}
\tau_L^{FE1} & = \frac{1}{|c| \, L},  \label{eqn:supg_rom_coeff_theo_1} \\
\tau_L^{FE2} & = \frac{1}{2|c| L} \Big(  coth(Pe) -\frac{1}{Pe} \Big), \quad Pe := \frac{|c|}{2\mu \, L}. \label{eqn:supg_rom_coeff_theo_2}
\end{align}
\end{subequations}

\begin{figure}[h!] 
\begin{center}
\includegraphics[width=0.3\textwidth,height=0.3\textwidth]{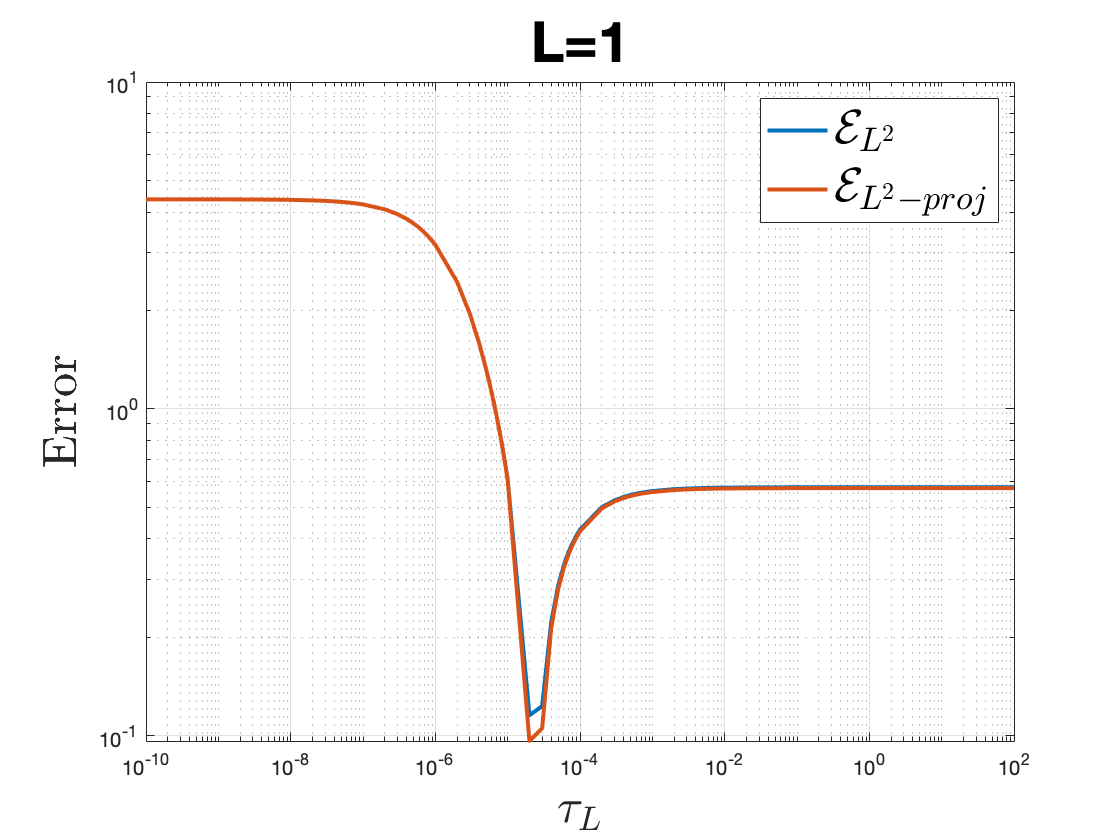}
\includegraphics[width=0.3\textwidth,height=0.3\textwidth]{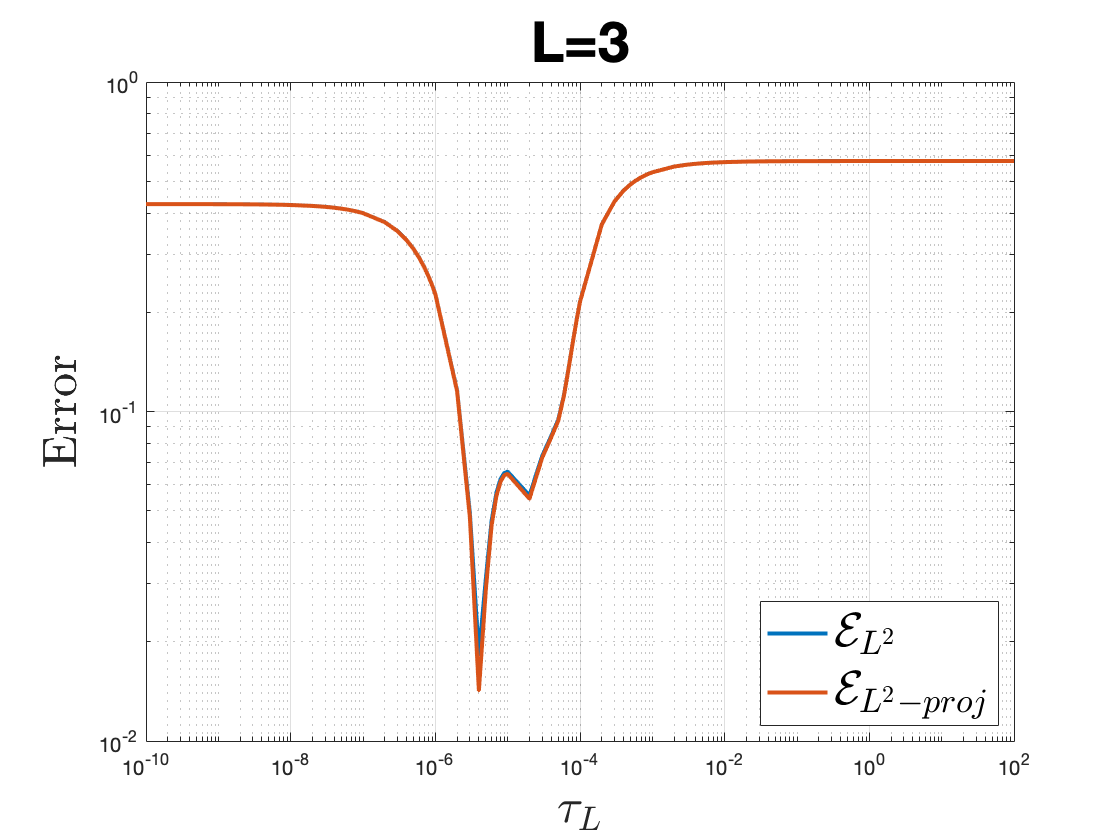}
\includegraphics[width=0.3\textwidth,height=0.3\textwidth]{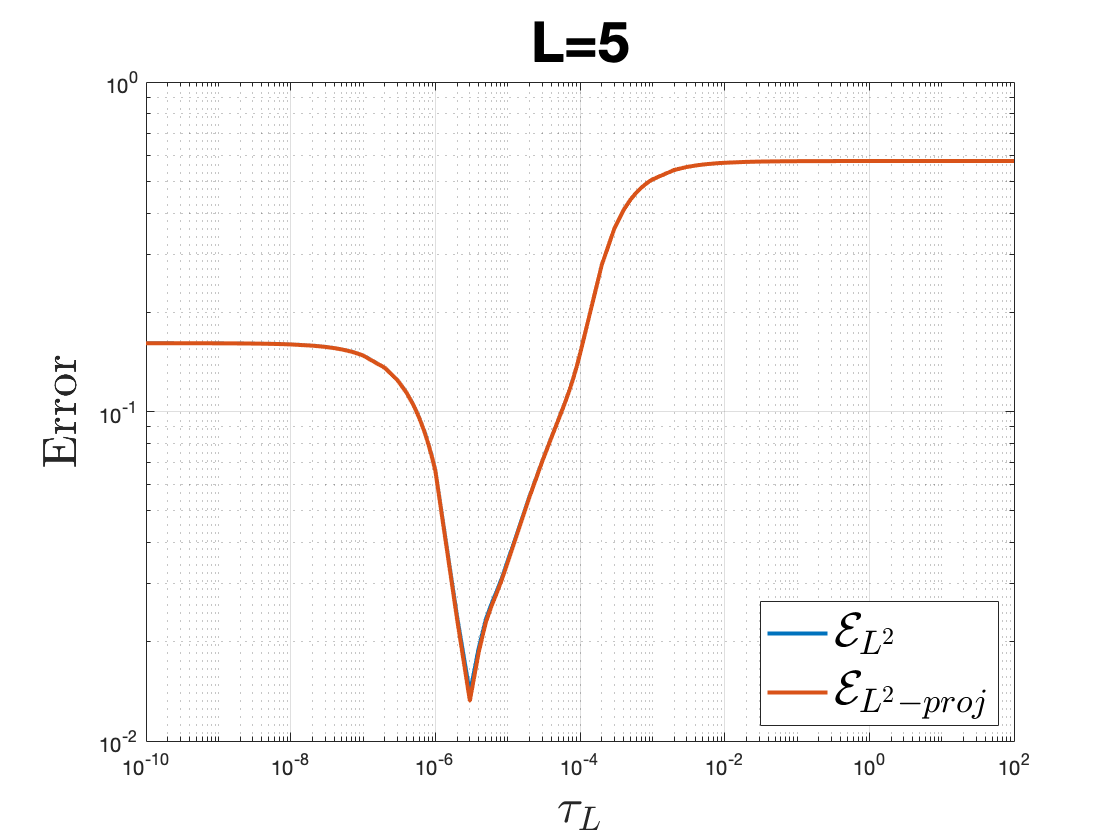}
\caption{$L^2$ error \eqref{eqn:l2_err} and $L^2$-projection error \eqref{eqn:l2_proj_err} for SUPG-ROM \eqref{eqn:matrix_vector_supg_rom} with $\mu^{testing}=0.5$ and different $L$ values for the stabilization coefficient $\tau_L$.}
\label{fig:trial_errror_supg_rom_coeff}
 \end{center} 
\end{figure}

The second strategy uses a trial and error approach to find the optimal stabilization coefficient $\tau_L$. We distinguish the trial and error approach in two ways, i.e., training and testing. In the training case, $\tau_L^{training}$ minimizes the difference between the ROM and FOM solutions over the whole training set parameters, i.e., $\mu^{training}$. In other words, the SUPG-ROM with $\tau_L^{training}$ minimizes the following quantity:
\begin{align}
\sum_{k=1}^{M_{training}} \Big( (\boldsymbol u_L(\tau_L^{training},\mu^{training}_k)- \boldsymbol u^{FOM}(\mu^{training}_k) \Big)^2, \label{eqn:min_trial_training}
\end{align}
where $M_{training}$ is the size of the training parameter set. In the testing case, $\tau_L^{testing}$ minimizes the difference between the ROM and FOM solutions just for the target parameter, i.e., $\mu^{testing}$. In other words, the SUPG-ROM with the stabilization coefficient $\tau_L^{testing}$ minimizes the following quantity:
\begin{align}
\Big( \boldsymbol u_L(\tau_L^{testing},\mu^{testing})- \boldsymbol u^{FOM}(\mu^{testing}) \Big)^2, \label{eqn:min_trial_testing}
\end{align}
where $\mu^{testing}$ falls outside the training range, as we will see in the next section. 
In Figure~\ref{fig:trial_errror_supg_rom_coeff}, we include the trial and error plots for different $L$ values. Specifically, we fix an interval for the stabilization coefficient $\tau_L$ and search for the optimal $\tau_L$ ($\tau^{testing}_L$) that minimizes the errors $\mathcal{E}_{L^2}$ defined in \eqref{eqn:l2_err} and $\mathcal{E}_{L^2-proj}$ defined in \eqref{eqn:l2_proj_err}.

In the third strategy, we use a data-driven approach to find the optimal stabilization coefficient $\tau_L^{d2}$ by creating the following ansatz:
\begin{align}
\boldsymbol A_{LS} \, \boldsymbol a_S \, \approx \,  \tau_L^{d2} \,  \boldsymbol A_{supg} \, \boldsymbol a_L,  \label{eqn:supg_d2_ansatz}
\end{align}
where $\tau_L^{d2}  \in \mathbbm{R}^{L \times L}$.

\begin{table}[h!] 
\centering
\begin{tabular}{| c| c| c| c|} 
\hline
& \textbf{Closure Term} &\textbf{Ansatz} & \textbf{Model}  \\ 
\hline 
\eqref{eqn:residual_based_ansatz1a}-\eqref{eqn:r1a_d2_vms_rom} & & & \\
Section~\ref{sec:r_d2_vms_rom} & $\boldsymbol a_S$ &  $\textbf{Ansatz}(Res_{\boldsymbol S}(\boldsymbol a_L)) = \widetilde{\boldsymbol A} \, Res_{\boldsymbol S}(\boldsymbol a_L)$ & \textbf{R1a-ROM}  \\ 
\hline 
\eqref{eqn:residual_based_ansatz1b}-\eqref{eqn:r1b_d2_vms_rom} & & & \\
Section~\ref{sec:r_d2_vms_rom} & $\boldsymbol a_S$ &   $\textbf{Ansatz}(Res_{\boldsymbol S}(\boldsymbol a_L)) = \widetilde{\boldsymbol A} \, Res_{\boldsymbol S}(\boldsymbol a_L) + \widetilde{ \boldsymbol b}$ & \textbf{R1b-ROM}  \\ 
\hline 
\hline
\eqref{eqn:residual_based_ansatz2_1a}-\eqref{eqn:r2_1a_d2_vms_rom} & $\boldsymbol a_S$ & $\textbf{Ansatz1}(Res_{\boldsymbol S}(\boldsymbol a_L)) = \widetilde{\boldsymbol A}_1 \, Res_{\boldsymbol S}(\boldsymbol a_L)$ & \\
Section~\ref{sec:r_d2_vms_rom} & $Res_{\boldsymbol L}( \boldsymbol a_S)$ &  $\textbf{Ansatz2}(Res_{\boldsymbol L}(\boldsymbol a_S)) =  \widetilde{\boldsymbol A}_2 \, Res_{\boldsymbol L}( \boldsymbol a_S )$ & \textbf{R2a-ROM}  \\ 
\hline 
\eqref{eqn:residual_based_ansatz2_2a}-\eqref{eqn:r2_2a_d2_vms_rom} & $\boldsymbol a_S$ & $\textbf{Ansatz1}(Res_{\boldsymbol S}(\boldsymbol a_L)) = \widetilde{\boldsymbol A}_1 \, Res_{\boldsymbol S}(\boldsymbol a_L) + \widetilde{ \boldsymbol b_1}$ & \\
Section~\ref{sec:r_d2_vms_rom} & $Res_{\boldsymbol L}( \boldsymbol a_S)$ &  $\textbf{Ansatz2}(Res_{\boldsymbol L}(\boldsymbol a_S)) =  \widetilde{\boldsymbol A}_2 \, Res_{\boldsymbol L}( \boldsymbol a_S ) + \widetilde{ \boldsymbol b_2}$ & \textbf{R2b-ROM}  \\ 
\hline 
\hline
\eqref{eqn:coeff_based_ansatz1a}-\eqref{eqn:c1a_d2_vms_rom} & & & \\
Section~\ref{sec:c_d2_vms_rom} & $\boldsymbol A_{LS} \boldsymbol a_S$ &   $\textbf{Ansatz}(\boldsymbol a_L) =  \widetilde{\boldsymbol A} \boldsymbol a_L$ & \textbf{C1a-ROM}  \\ 
\hline
\eqref{eqn:coeff_based_ansatz1b}-\eqref{eqn:c1b_d2_vms_rom} & & & \\
Section~\ref{sec:c_d2_vms_rom} & $\boldsymbol A_{LS} \boldsymbol a_S$ &   $\textbf{Ansatz}(\boldsymbol a_L) =  \widetilde{\boldsymbol A} \boldsymbol a_L + \widetilde{ \boldsymbol b}$ & \textbf{C1b-ROM}  \\ 
\hline 
\end{tabular}
\caption{Summary of the numerical methods based on the D2-VMS-ROM.}
 \label{table:d2vmsrom_methods}
\end{table}

\begin{table}[h!] 
\centering
\begin{tabular}{|c|c|c|} 
\hline
& \textbf{Stabilization Coefficient} & \textbf{Model}  \\ 
\hline
\eqref{eqn:matrix_vector_supg_rom}, \eqref{eqn:supg_rom_coeff_theo_1}-\eqref{eqn:supg_rom_coeff_theo_2} in Section~\ref{sec:supg_rom} & $\tau_L^{FE1}$, $\tau_L^{FE2}$  & \textbf{SUPG-ROM}($\tau_L^{FE1}$,$\tau_L^{FE2}$) \\ 
\hline
\eqref{eqn:matrix_vector_supg_rom}, \eqref{eqn:min_trial_training} in Section~\ref{sec:supg_rom} & $\tau_L^{training}$ & \textbf{SUPG-ROM}($\tau_L^{training}$)  \\ 
\hline
\eqref{eqn:matrix_vector_supg_rom}, \eqref{eqn:min_trial_testing} in Section~\ref{sec:supg_rom} & $\tau_L^{testing}$ & \textbf{SUPG-ROM}($\tau_L^{testing}$)   \\ 
\hline
\eqref{eqn:matrix_vector_supg_rom}, \eqref{eqn:supg_d2_ansatz} in Section~\ref{sec:supg_rom} & $\tau_L^{d2}$ & \textbf{SUPG-ROM}($\tau_L^{d2}$)   \\ 
\hline
\end{tabular}
\caption{Summary of the numerical methods based on the SUPG-ROM \eqref{eqn:matrix_vector_supg_rom}.}
 \label{table:supgrom_methods}
\end{table}

%%%%%%%%%%%%%
\section{Numerical Results} \label{sec:numerical_results}
In this section, we investigate the numerical accuracy of the R-D2-VMS-ROMs in Section~\ref{sec:r_d2_vms_rom}, C-D2-VMS-ROMs in Section~\ref{sec:c_d2_vms_rom}, and SUPG-ROMs in Section~\ref{sec:supg_rom}. Furthermore, we compare the consistency behavior of the R-D2-VMS-ROMs and C-D2-VMS-ROMs.
For clarity, in Tables~\ref{table:d2vmsrom_methods} and \ref{table:supgrom_methods}, we include a summary of the different D2-VMS-ROMs and SUPG-ROMs investigated in this section, respectively. 

We present numerical results for the parameter-dependent CD problem~\eqref{eqn:parameter_cd_problem} with the following exact solution, which was used in \cite{rebollo2015variational}:
\begin{align} \label{eqn:exact_soln}
u(x,\mu) = \frac{\exp(c x/\mu )-1}{\exp(c/\mu)-1} - x, 
\end{align}
where $\mu$ is the diffusion coefficient. The forcing term is $f = -c = - 400$. 

\paragraph{Snapshot Generation}
We generate the FOM results for $\mu \in [1,10]$ with $\Delta \mu = 1 $ values by using a linear FE spatial discretization with mesh size $h=1/4096$.

\paragraph{ROM Construction}
We generate the ROM basis functions and operators by collecting the snapshots, which are the solutions of problem \eqref{eqn:parameter_cd_problem} for $\mu^{training}=1,2,...,9,10$. 
To build the ROM basis, we use the POD~\cite{HLB96,volkwein2013proper}.
To train the D2-VMS-ROM operators, $\widetilde{\boldsymbol A}$ and  $\widetilde{\boldsymbol b}$, we use the FOM data for all the $\mu^{training}$ values. We test all the ROMs for $\mu^{testing} = 0.5, 0.1, 0.05, 15$, which fall outside the training range. Thus, we are testing all the ROMs in the \textit{predictive regime}.

To compare the numerical accuracy of the methods defined in Section~\ref{sec:d2_vms_rom} and \ref{sec:supg_rom}, we use different metrics. In Section~\ref{sec:numerical_results_single_mu}, for one testing parameter value, we use the $L^2$ error \eqref{eqn:l2_err} and $L^2$-projection error \eqref{eqn:l2_proj_err}, which measure the difference between the ROM solution $\boldsymbol u_L$ and FOM solution $\boldsymbol u^{FOM}$, and between the ROM solution $\boldsymbol u_L$ and the $L^2$-projection of the FOM solution, respectively. In Section~\ref{sec:all_mu}, we use avg-$L^2$ error \eqref{eqn:avg_l2_err} for all testing parameters.

%%%%
\subsection{Testing the ROMs for $\mu^{testing} = 0.5$}   \label{sec:numerical_results_single_mu}
In this section, we compare the accuracy of all the ROMs considering the following metrics:
\begin{subequations}
\begin{align}
\mathcal{E}_{L^2} &= \left\| \, \boldsymbol u_L(\mu^{testing},operators) \, - \, \boldsymbol  u^{FOM}(\mu^{testing}) \,\right\|_{L^2},  \label{eqn:l2_err} \\
\mathcal{E}_{L^2-proj} &= \left\| \, \boldsymbol u_L(\mu^{testing},operators) \, - \, \sum_{i=1}^{L} \Big\langle\boldsymbol u^{FOM}(\mu^{testing}), \boldsymbol \varphi_i \Big\rangle  \, \boldsymbol \varphi_i \,  \right\|_{L^2},  \label{eqn:l2_proj_err}
\end{align}
\end{subequations}
where $\boldsymbol u_L$ and $\boldsymbol u^{FOM}$ are the ROM and FOM solutions, and $operators$ are the D2 operators from Section~\ref{sec:d2_vms_rom} or the stabilization term \eqref{eqn:matrix_vector_supg_rom} from Section~\ref{sec:supg_rom}. In this section, the metrics~\eqref{eqn:l2_err} and \eqref{eqn:l2_proj_err} are computed for just one testing parameter value, $\mu^{testing}=0.5$, which is outside the training range.

In Tables~\ref{table:rom_err_m1_t05} and \ref{table:rom_err_m2_t05}, we list the ROM errors for all the methods that were introduced in Sections~\ref{sec:c_d2_vms_rom} and \ref{sec:r_d2_vms_rom}, which are run for the testing parameter $\mu^{testing}=0.5$ (outside the training range) and assessed by using the metrics \eqref{eqn:l2_err}-\eqref{eqn:l2_proj_err}.

\begin{table}[h!] 
\centering
\begin{tabular}{|c|c|c|c|c|c|c|c|c|c|} 
\hline
$L$ & G-ROM & R1a-ROM & R1b-ROM & R2a-ROM & R2b-ROM & C1a-ROM & C1b-ROM \\ 
\hline
1 & 4.38e+00 & 6.37e-02 & 6.37e-02 & 6.37e-02 & 6.37e-02 & 5.72e+00 & 6.38e-02 \\
2 & 3.72e-01 & 2.49e-02 & 2.49e-02 & 2.49e-02 & 2.49e-02 & 3.66e-01 & 2.51e-02 \\
3 & 4.26e-01 & 1.15e-02 & 1.15e-02 & 1.15e-02 & 1.15e-02 & 4.19e-01 & 1.16e-02 \\
4 & 1.68e-01 & 6.55e-03 & 6.55e-03 & 6.55e-03 & 6.55e-03 & 1.68e-01 & 6.69e-03 \\
5 & 1.61e-01 & 4.45e-03 & 4.45e-03 & 4.45e-03 & 4.45e-03 & 1.61e-01 & 4.56e-03 \\
6 & 9.92e-02 & 3.33e-03 & 3.33e-03 & 3.33e-03 & 3.33e-03 & 9.92e-02 & 3.41e-03 \\
7 & 9.49e-02 & 2.63e-03 & 2.63e-03 & 2.63e-03 & 2.63e-03 & 9.49e-02 & 2.69e-03 \\
\hline
\end{tabular}
\caption{$L^2$ error \eqref{eqn:l2_err} for G-ROM, R1-ROMs, R2-ROMs, and C1-ROMs for various $L$ values (number of POD modes).} \label{table:rom_err_m1_t05}
\end{table}

\begin{table}[h!] 
\centering
\begin{tabular}{|c|c|c|c|c|c|c|c|c|} 
\hline
$L$ & G-ROM & R1a-ROM & R1b-ROM & R2a-ROM & R2b-ROM & C1a-ROM & C1b-ROM \\ 
\hline
1 & 4.38e+00 & 5.45e-04 & 5.45e-04 & 5.46e-04 & 5.46e-04 & 5.72e+00 & 2.84e-03 \\
2 & 3.71e-01 & 1.04e-04 & 9.29e-05 & 1.04e-04 & 1.04e-04 & 3.66e-01 & 3.53e-03 \\
3 & 4.26e-01 & 6.90e-05 & 7.22e-05 & 6.98e-05 & 7.21e-05  & 4.19e-01 & 2.06e-03 \\
4 & 1.68e-01 & 8.20e-05 & 8.18e-05 & 8.10e-05 & 8.07e-05 & 1.68e-01 & 1.38e-03 \\
5 & 1.61e-01 & 2.48e-04 & 2.47e-04 & 2.13e-04 & 2.13e-04 & 1.61e-01 & 1.00e-03 \\
6 & 9.92e-02 & 1.72e-04 & 1.72e-04 & 1.39e-04 & 1.39e-04 & 9.92e-02 & 7.59e-04 \\
7 & 9.48e-02 & 1.18e-04 & 1.18e-04 & 1.44e-04 & 1.44e-04 & 9.48e-02 & 5.95e-04 \\
\hline
\end{tabular}
\caption{$L^2$-projection error \eqref{eqn:l2_proj_err} for G-ROM, R1-ROMs, R2-ROMs,and C1-ROMs for various $L$ values (number of POD modes).} \label{table:rom_err_m2_t05}
\end{table} 

For both Tables~\ref{table:rom_err_m1_t05} and \ref{table:rom_err_m2_t05}, the G-ROM and C1a-ROM yield the worst results among all the ROMs. Furthermore, the R1a, R1b, R2a, and R2b-ROMs give the lowest errors among all the ROMs. In Tables~\ref{table:rom_err_m1_t05} and \ref{table:rom_err_m2_t05}, the R1a and R1b, and R2a and R2b-ROM errors are similar, and almost coincide when $L$ approaches $d$. 
Thus, these numerical results suggest that there is no benefit in adding the term $\widetilde{\boldsymbol b}$ to the residual-based ansatzes. For this reason, to list the errors, starting with the next section, we use only R1a-ROM and R2a-ROM from the four R-D2-VMS-ROMs. In contrast, for the comparison of the coefficient-based data-driven models, we observe that C1b-ROM yields more accurate results than C1a-ROM.
Thus, to make a fair comparison among all ROMs in the next section, we list C1b-ROM since it is the most accurate coefficient-based model.

We note that the errors in Table~\ref{table:rom_err_m2_t05} are equal to or significantly lower than the errors in Table~\ref{table:rom_err_m1_t05}. This is due to the fact that, in this case, the projection error 
\begin{equation} \label{eqn:proj_err}
\left\| \, \boldsymbol u^{FOM}(\mu^{testing}) \, - \, \sum_{i=1}^{L} \Big\langle\boldsymbol u^{FOM}(\mu^{testing}), \boldsymbol \varphi_i \Big\rangle  \, \boldsymbol \varphi_i \,  \right\|_{L^2}
\end{equation}
dominates the error bound for \eqref{eqn:l2_err}. In view of the smallest errors achieved with the metric \eqref{eqn:l2_proj_err}, to assess the ROM accuracy starting with the next section, we use only that metric.

To have a better understanding of the advantage of using the residual-based ansatz over the coefficient-based ansatz to model the closure term, in Tables~\ref{table:c1_rom_consistency_table_mu05}, \ref{table:r1_rom_consistency_table_mu05}, and \ref{table:r2_rom_consistency_table_mu05}, we investigate the ROM consistency, by listing the $L^2$-norm of the closure term and its ansatz for each ROMs. In our numerical investigation, we call a closure model consistent if the magnitude of its ansatz is on the same order as the magnitude of the closure term.

In Table~\ref{table:c1_rom_consistency_table_mu05}, we investigate the consistency behavior of the C-D2-VMS-ROMs, i.e., C1a-ROM~\eqref{eqn:c1a_d2_vms_rom} and C1b-ROM~\eqref{eqn:c1b_d2_vms_rom} by listing
$\|\boldsymbol A_{LS} \, \boldsymbol a_S\|_{L^2}$, which is the norm of the closure term (which is the same for both models), and $\|\widetilde{\boldsymbol A} \boldsymbol a_L\|_{L^2}$ and $\|\widetilde{\boldsymbol A} \boldsymbol a_L + \widetilde{ \boldsymbol b}\|_{L^2}$, which are the norms of the C1a-ROM and C1b-ROM ansatzes, respectively, for all $L$ values.
Based on the numerical results, we observe that the norm of the C1b-ROM ansatz, $\|\widetilde{\boldsymbol A} \boldsymbol a_L + \widetilde{ \boldsymbol b}\|_{L^2}$, has the same order of magnitude as the norm of the closure term, $\|\boldsymbol A_{LS} \, \boldsymbol a_S\|_{L^2}$. In contrast, the order of magnitude of the norm of the C1a-ROM ansatz, $\|\widetilde{\boldsymbol A} \boldsymbol a_L\|_{L^2}$, quickly diminishes. Thus, we conclude that the C1b-ROM consistency error is smaller than the C1a-ROM consistency error.

We perform the same consistency investigation for the R1-D2-VMS-ROMs and R2-D2-VMS-ROMs. In Table~\ref{table:r1_rom_consistency_table_mu05}, we list the $L^2$-norm of the closure term, $\|\boldsymbol a_S\|_{L^2}$, the norm of the the R1a-ROM ansatz, $\|\widetilde{\boldsymbol A} \, Res_{\boldsymbol S}(\boldsymbol a_L) \|_{L^2}$, and the norm of the R1b-ROM ansatz, $\|\widetilde{\boldsymbol A} \, Res_{\boldsymbol S}(\boldsymbol a_L) + \widetilde{ \boldsymbol b}\|_{L^2}$, for all $L$ values. 
Likewise, in Table~\ref{table:r2_rom_consistency_table_mu05}, we list the $L^2$-norm of the closure term, $\|Res_{\boldsymbol L}(\boldsymbol a_S)\|_{L^2}$, the norm of the the R2a-ROM ansatz $\|\widetilde{\boldsymbol A}_2 \, Res_{\boldsymbol L}( \boldsymbol a_S^{approx} )\|_{L^2}$, and the norm of the R2b-ROM ansatz, $\|\widetilde{\boldsymbol A}_2 \, Res_{\boldsymbol L}( \boldsymbol a_S^{approx} ) + \widetilde{ \boldsymbol b_2}\|_{L^2}$, for all $L$ values.
In both tables, we observe that the orders of the magnitude of the norms of the R1-ROM and R2-ROM ansatzes are the same as their closure terms. Furthermore, as $L$ goes to $d$, the norm of the R1a and R1b, and R2a and R2b-ROM ansatzes yield the same results. Thus, we conclude that R1a-ROM, R1b-ROM, R2a-ROM, and R2b-ROM have similar small consistency errors.

\begin{remark} \label{remark:d2_vms_rom}
From Tables~\ref{table:rom_err_m1_t05}-\ref{table:rom_err_m2_t05}, i.e., the accuracy tables, and Tables~\ref{table:c1_rom_consistency_table_mu05}-\ref{table:r2_rom_consistency_table_mu05}, i.e., the consistency tables, we observe that R1a-ROM and R1b-ROM, and R2a-ROM and R2b-ROM yield similar results; thus, in next section, we only keep R1a-ROM and R2a-ROM. Furthermore, C1b-ROM is more accurate and has a smaller consistency error than C1a-ROM.
Thus, to have a fair comparison, in the next section, we focus our investigation on C1b-ROM. Furthermore, to better assess the accuracy, we use metric \eqref{eqn:l2_proj_err}.

In what follows, for ease of notation, we denote C1b-ROM, R1a-ROM, and R2a-ROM as C1-ROM, R1-ROM, and R2-ROM, respectively.
\end{remark}

\begin{table}[h!] 
\centering
\begin{tabular}{|c|c|c|c|} 
\hline
 & \text{Closure Term}  & \text{Ansatz in} C1a  & \text{Ansatz in} C1b  \\
\hline
$L$ &  $\|\boldsymbol A_{LS} \, \boldsymbol a_S\|_{L^2}$  &  $\|\widetilde{\boldsymbol A} \boldsymbol a_L\|_{L^2}$ & $\|\widetilde{\boldsymbol A} \boldsymbol a_L + \widetilde{ \boldsymbol b}\|_{L^2}$  \\
\hline
1 & 3.06e+02 & 7.10e+02 & 3.08e+02 \\
2 & 1.21e+03 & 1.77e+01 & 1.22e+03 \\
3 & 1.21e+03 & 1.90e+01 & 1.26e+03 \\
4 & 1.00e+03 & 7.07e-01 & 1.12e+03 \\
5 & 7.85e+02 & 3.77e-01 & 9.85e+02 \\
6 & 5.90e+02 & 2.41e-02 & 8.70e+02 \\
7 & 4.18e+02 & 4.60e-04 & 7.77e+02 \\
8 & 2.64e+02 & 3.32e-04 & 7.00e+02 \\
9 & 1.26e+02 & 2.42e-02 & 6.36e+02 \\
10 & 0 & 0 & 0 \\
\hline 
\end{tabular}
\caption{Consistency error comparison for C-D2-VMS-ROMs, i.e., C1a-ROM and C1b-ROM, for various $L$ values (number of POD modes).} \label{table:c1_rom_consistency_table_mu05}
\end{table}

\begin{table}[h!] 
\centering
\begin{tabular}{|c|c|c|c|} 
\hline
& \text{Closure Term} & \text{Ansatz in} R1a  & \text{Ansatz in} R1b   \\
\hline
$L$ &  $\|\boldsymbol a_S\|_{L^2}$  &  $\|\widetilde{\boldsymbol A} \, Res_{\boldsymbol S}(\boldsymbol a_L)\|_{L^2}$ & $\|\widetilde{\boldsymbol A} \, Res_{\boldsymbol S}(\boldsymbol a_L) + \widetilde{ \boldsymbol b}\|_{L^2}$ \\
\hline
1 & 6.37e-02 & 7.01e-02 & 7.01e-02 \\
2 & 2.48e-02 & 2.67e-02 & 2.66e-02 \\
3 & 1.14e-02 & 1.28e-02 & 1.28e-02 \\
4 & 6.36e-03 & 8.77e-03 & 8.77e-03 \\
5 & 4.17e-03 & 7.12e-03 & 7.11e-03 \\
6 & 2.95e-03 & 6.03e-03 & 6.03e-03 \\
7 & 2.13e-03 & 5.12e-03 & 5.12e-03 \\
8 & 1.50e-03 & 7.41e-04 & 7.41e-04 \\
9 & 9.35e-04 & 4.23e-04 & 4.23e-04 \\
10 & 0 & 0 & 0 \\
\hline 
\end{tabular}
\caption{Consistency error comparison for R1-D2-VMS-ROMs, i.e., R1a-ROM and R1b-ROM, for various $L$ values (number of POD modes).} \label{table:r1_rom_consistency_table_mu05}
\end{table}

\begin{table}[h!] 
\centering
\begin{tabular}{|c|c|c|c|} 
\hline
& \text{Closure Term} & \text{Ansatz in} R2a & \text{Ansatz in} R2b \\
\hline
$L$ & $\|Res_{\boldsymbol L}( \boldsymbol a_S)\|_{L^2}$  &  $\|\widetilde{\boldsymbol A}_2 \, Res_{\boldsymbol L}( \boldsymbol a_S^{approx})\|_{L^2}$ & $\|\widetilde{\boldsymbol A}_2 \, Res_{\boldsymbol L}( \boldsymbol a_S^{approx} ) + \widetilde{ \boldsymbol b_2}\|_{L^2}$ \\
\hline
1 & 4.24e+01 & 4.02e+01 & 4.25e+01 \\
2 & 1.24e+03 & 1.27e+03 & 6.20e+02 \\
3 & 1.21e+03 & 1.31e+03 & 1.89e+04 \\
4 & 1.06e+03 & 1.23e+03 & 1.23e+03 \\
5 & 8.34e+02 & 1.10e+03 & 1.10e+03 \\
6 & 6.86e+02 & 1.02e+03 & 1.02e+03 \\
7 & 5.31e+02 & 9.17e+02 & 9.17e+02 \\
8 & 4.43e+02 & 7.92e+02 & 7.92e+02 \\
9 & 3.74e+02 & 6.90e+02 & 6.87e+02 \\
10 & 0 & 0 & 0 \\
\hline 
\end{tabular}
\caption{Consistency error comparison for R2-D2-VMS-ROMs, i.e., R2a-ROM and R2b-ROM, for various $L$ values (number of POD modes).} \label{table:r2_rom_consistency_table_mu05}
\end{table}

%%%%%%%%%%%%%supg-rom 
Until now, we provided numerical results for the D2-VMS-ROMs. To ensure a fair comparison, we present hereafter the SUPG-ROM numerical results. 
In Tables~\ref{table:supg_rom_err_m1_t05} and \ref{table:supg_rom_err_m2_mu05}, we list the SUPG-ROM errors by using different stabilization coefficients $\tau_L$ and the metrics \eqref{eqn:l2_err}-\eqref{eqn:l2_proj_err}. 
For both Tables~\ref{table:supg_rom_err_m1_t05} and \ref{table:supg_rom_err_m2_mu05}, the second and third columns, i.e., $\text{SUPG}(\tau_L^{FE1})$ and $\text{SUPG}(\tau_L^{FE2})$, where $\tau_L^{FE1}$ and $\tau_L^{FE2}$ are calculated by using \eqref{eqn:supg_rom_coeff_theo_1}-\eqref{eqn:supg_rom_coeff_theo_2}, give the highest SUPG-ROM errors. 
The fourth and fifth columns show $\text{SUPG}(\tau_L^{training})$ and $\text{SUPG}(\tau_L^{testing})$ errors, where $\tau_L^{training}$ and $\tau_L^{testing}$ are calculated by using \eqref{eqn:min_trial_training}-\eqref{eqn:min_trial_testing}. In the last column, the SUPG-ROM errors are calculated by modeling the stabilization coefficient $\tau_L$ with a data-driven approach, see \eqref{eqn:supg_d2_ansatz}. 
Among all SUPG-ROMs with different $\tau_L$ models in Tables~\ref{table:supg_rom_err_m1_t05} and \ref{table:supg_rom_err_m2_mu05}, the $\text{SUPG}(\tau_L^{testing})$ yields the lowest error with respect to both metrics \eqref{eqn:l2_err}-\eqref{eqn:l2_proj_err}. Thus, in the next section, we compute the SUPG-ROM errors only considering the stabilization coefficient $\tau^{testing}_L$~\eqref{eqn:min_trial_testing}.

Furthermore, the errors in Table~\ref{table:supg_rom_err_m2_mu05} are similar to the errors in Table~\ref{table:supg_rom_err_m1_t05}. This is due to the fact that, in this case, the projection error \eqref{eqn:proj_err} does not dominate the error bound in \eqref{eqn:l2_err} so the metrics \eqref{eqn:l2_err} and \eqref{eqn:l2_proj_err} yield similar results. Thus, in the next section, we use metric \eqref{eqn:l2_proj_err} to calculate the SUPG-ROM errors. 

\begin{remark} \label{remark:supg_rom}
From Tables~\ref{table:supg_rom_err_m1_t05} and \ref{table:supg_rom_err_m2_mu05}
, we observe that we have the lowest SUPG-ROM error when the stabilization coefficient $\tau_L$ is chosen as $\tau^{testing}_L$~\eqref{eqn:min_trial_testing} and metric \eqref{eqn:l2_proj_err} is used. In the next section, we only consider the $\text{SUPG}(\tau_L^{testing})$ with the metric \eqref{eqn:l2_proj_err}.
\end{remark}

\begin{table}[h!] 
\centering
\begin{tabular}{|c|c|c|c|c|c| } 
\hline
$L$ &  \text{SUPG}($\tau_L^{FE1}$) &  \text{SUPG}($\tau_L^{FE2}$) &  \text{SUPG}($\tau_L^{training}$) &  \text{SUPG}($\tau_L^{testing}$) & \text{SUPG}($\tau_L^{d2}$) \\
\hline
1 &  5.70e-01 & 5.63e-01 & 1.23e-01 & 1.16e-01 & 5.72e+00 \\
2 &  5.51e-01 & 5.27e-01 & 8.35e-02 & 3.11e-02 & 4.62e-01 \\ 
3 &  5.23e-01 & 4.72e-01 & 5.56e-02 & 1.83e-02 & 5.10e-01 \\ 
4 &  4.86e-01 & 4.00e-01 & 7.06e-02 & 1.59e-02 & 1.71e-01 \\ 
5 &  4.40e-01 & 3.22e-01 & 7.06e-02 & 1.40e-02 & 1.62e-01 \\ 
6 &  3.89e-01 & 2.53e-01 & 7.05e-02 & 1.40e-02 & 9.92e-02 \\ 
7 &  3.38e-01 & 2.05e-01 & 7.05e-02 & 1.36e-02 & 9.49e-02 \\ 
\hline 
\end{tabular}
\caption{$L^2$ error \eqref{eqn:l2_err} for SUPG-ROM \eqref{eqn:matrix_vector_supg_rom} for various $\tau_L$ models and $L$ values (number of POD modes).} \label{table:supg_rom_err_m1_t05}
\end{table}

\begin{table}[h!] 
\centering
\begin{tabular}{|c|c|c|c|c|c| } 
\hline
$L$ &  \text{SUPG}($\tau_L^{FE1}$) &  \text{SUPG}($\tau_L^{FE2}$) &  \text{SUPG}($\tau_L^{training}$) &  \text{SUPG}($\tau_L^{testing}$) & \text{SUPG}($\tau_L^{d2}$) \\
\hline
1 &  5.66e-01 & 5.60e-01 & 1.05e-01 & 9.64e-02 & 5.72e+00 \\
2 &  5.51e-01 & 5.26e-01 & 7.97e-02 & 1.87e-02 & 4.61e-01 \\ 
3 &  5.23e-01 & 4.72e-01 & 5.44e-02 & 1.42e-02 & 5.10e-01 \\ 
4 &  4.86e-01 & 4.00e-01 & 7.02e-02 & 1.45e-02 & 1.71e-01 \\ 
5 &  4.40e-01 & 3.22e-01 & 7.05e-02 & 1.33e-02 & 1.61e-01 \\ 
6 &  3.89e-01 & 2.53e-01 & 7.04e-02 & 1.36e-02 & 9.92e-02 \\ 
7 &  3.38e-01 & 2.05e-01 & 7.04e-02 & 1.33e-02 & 9.48e-02 \\ 
\hline 
\end{tabular}
\caption{$L^2$-projection error \eqref{eqn:l2_proj_err} for SUPG-ROM \eqref{eqn:matrix_vector_supg_rom} for various $\tau_L$ models and $L$ values (number of POD modes).} \label{table:supg_rom_err_m2_mu05}
\end{table}

\subsection{Testing for all $\mu$ values}  \label{sec:all_mu}
In this section, instead of further investigating different single $\mu^{testing}$ values that are out of the training set (as we did in Section~\ref{sec:numerical_results_single_mu}), we use the average $L^2$ error~\eqref{eqn:avg_l2_err}, which is the average of the metric \eqref{eqn:l2_proj_err}, to measure the  ROM accuracy for several parameter values that are out of the training set, $\mu^{testing} = 0.5, 0.1, 0.05, 15$:

\begin{align}
\mathcal{E}_{avg-L^2} &= \frac{1}{M} \sum_{k=1}^{M} \left\| \, \boldsymbol u_L(\mu_k^{testing},operators) \, - \, \sum_{i=1}^{r} \Big\langle\boldsymbol u^{FOM}(\mu_k^{testing}), \boldsymbol \varphi_i \Big\rangle  \, \boldsymbol \varphi_i \,  \right\|_{L^2}.  \label{eqn:avg_l2_err}
\end{align}
In Table~\ref{table:rom_err_avg}, we list the average $L^2$ errors for G-ROM, C1-ROM , R1-ROM, R2-ROM, and SUPG-ROM with $\tau_L^{testing}$ model (see Tables~\ref{table:d2vmsrom_methods}-\ref{table:supgrom_methods} and 
Remarks~\ref{remark:d2_vms_rom}-\ref{remark:supg_rom} for summary of models).

\begin{table}[h!] 
\centering
\begin{tabular}{|c|c|c|c|c|c|c|c|} 
\hline
$L$ & G-ROM & R1-ROM & R2-ROM & C1-ROM & \text{SUPG}($\tau_L^{testing}$)  \\ 
\hline
1 & 1.95e+01 & 1.10e-03 & 1.10e-03 & 3.99e-03 & 1.47e-01 \\
2 & 4.17e-01 & 3.11e-04 & 3.11e-04 & 5.76e-03 & 7.07e-02 \\
3 & 3.31e+00 & 2.85e-04 & 2.86e-04 & 5.06e-03 & 4.57e-02 \\
4 & 3.07e-01 & 3.36e-04 & 3.30e-04 & 4.68e-03 & 3.11e-02 \\
5 & 1.93e+00 & 1.03e-03 & 7.26e-04 & 4.43e-03 & 2.30e-02 \\
6 & 2.64e-01 & 8.30e-04 & 5.14e-04 & 4.17e-03 & 1.90e-02 \\
7 & 1.49e+00 & 6.54e-04 & 8.41e-04 & 3.95e-03 & 1.67e-02 \\
\hline
\end{tabular}
\caption{Average $L^2$ error \eqref{eqn:avg_l2_err} for G-ROM, C1-ROM, R1-ROM, R2-ROM, and $\text{SUPG}(\tau_L^{testing})$ for various $L$ values (number of POD modes).} \label{table:rom_err_avg}
\end{table}

Based on the errors in Table~\ref{table:rom_err_avg}, G-ROM yields the least accurate results. SUPG-ROM is more accurate than G-ROM and less accurate than C1-ROM, R1-ROM, and R2-ROM. This is not surprising, as SUPG-ROM has fewer degrees of freedom than the D2-VMS-ROMs to be adjusted in the offline step. The R1-ROM and R2-ROM errors yield much better accuracy than the C1-ROM. For example, with $L=2,3,4$, the R1-ROM and R2-ROM are more than one order of magnitude more accurate than the C1-ROM. This conclusion is the main result of this paper, which points out that the consistent models R1-ROM and R2-ROM, which are residual-based models, yield significantly better accuracy than the coefficient-based C1-ROM.
Finally, although R2-ROM involves more information from the sub-scale equation~\eqref{eqn:matrix_vector_form_small},  R1-ROM and R2-ROM yield similar results, as we observed in Tables~\ref{table:rom_err_m1_t05} and \ref{table:rom_err_m2_t05}.

\section{Conclusions and Outlook} \label{sec:conclusions_outlook}
In this paper, we proposed and investigated data-driven ROM closure modeling for convection-dominated problems. Specifically, we considered three types of ROM closure modeling using available data to construct accurate ROM closure operators. 
The first type of data-driven ROM closure was a novel residual-based D2-VMS-ROM (R-D2-VMS-ROM), in which the closure term is a function of the ROM residual.
The second type of data-driven ROM closure that we investigated was the standard coefficient-based D2-VMS-ROM (C-D2-VMS-ROM)~\cite{mou2021data,xie2018data}. 
We also investigated a SUPG-ROM stabilization strategy in which the stabilization parameter was computed by extrapolating the classical FE scalings in the ROM framework, by a trial and error approach, or by using data-driven modeling.
Finally, for comparison purposes, we investigated a standard G-ROM in which neither stabilization nor closure was used. We investigated the new R-D2-VMS-ROM and the standard C-D2-VMS-ROM, SUPG-ROM, and G-ROM  in the numerical simulation of a one-dimensional parameter-dependent convection-diffusion problem in the predictive regime (i.e., for parameters that were not used in the training stage). 

Our numerical investigation yielded the following conclusions:
First, the novel residual-based D2-VMS-ROM, i.e., R-D2-VMS-ROM was the most accurate model.
Second, R-D2-VMS-ROM yielded smaller consistency errors (i.e., smaller error between the modeling ansatz and the true closure term) than the standard C-D2-VMS-ROM. Third, R-D2-VMS-ROM, C-D2-VMS-ROM, and SUPG-ROM were significantly more accurate than G-ROM. Finally, the SUPG-ROM with the stabilization coefficient $\tau^{testing}_L$~\eqref{eqn:min_trial_testing}, which corresponds to the trial and error framework, was more 
accurate than the SUPG-ROM with the FE extrapolated stabilization parameter and data-driven parameter.

There are several research directions that can be pursued next.
Probably the most important is the extension of the new R-D2-VMS-ROM framework to more complex convection-dominated problems, such as under-resolved turbulent flows.
Another important research direction is providing mathematical support for the new R-D2-VMS-ROM. For example, we plan to prove R-D2-VMS-ROM's verifiability, i.e., to show that when the closure model error decreases, the ROM error decreases at the same rate.  The first step in this direction has been taken in~\cite{koc2022verifiability}, where we proved the verifiability of the standard C-D2-VMS-ROM.

\paragraph{Acknowledgments:} 
The first and third authors are partially funded by Programma Operativo FEDER Andalucia 2014-2020 grant US-1254587 and ARIA MSCA-RISE EU Grant 872442. The second author is supported by Project PID2021-123153OB-C21 funded by MCIN/AEI/10.13039/501100011033/FEDER, UE and ARIA MSCA-RISE EU Grant 872442. The fourth author is funded by ARIA MSCA-RISE EU Grant 872442 and National Science Foundation grant DMS-2012253.

\bibliographystyle{abbrv}
\bibliography{reference}

\begin{thebibliography}{10}

\bibitem{ARCME}
N.~Ahmed, T.~Chac\'{o}n~Rebollo, V.~John, and S.~Rubino.
\newblock A review of variational multiscale methods for the simulation of
  turbulent incompressible flows.
\newblock {\em Arch. Comput. Methods Eng.}, 24(1):115--164, 2017.

\bibitem{ahmed2021closures}
S.~E. Ahmed, S.~Pawar, O.~San, A.~Rasheed, T.~Iliescu, and B.~R. Noack.
\newblock On closures for reduced order models $-$ a spectrum of
  first-principle to machine-learned avenues.
\newblock {\em Phys. Fluids}, 33(9):091301, 2021.

\bibitem{AzaiezJCP21}
M.~Aza\"{\i}ez, T.~Chac\'{o}n~Rebollo, and S.~Rubino.
\newblock A cure for instabilities due to advection-dominance in {POD} solution
  to advection-diffusion-reaction equations.
\newblock {\em J. Comput. Phys.}, 425:109916, 2021.

\bibitem{ballarin2015supremizer}
F.~Ballarin, A.~Manzoni, A.~Quarteroni, and G.~Rozza.
\newblock Supremizer stabilization of {POD--G}alerkin approximation of
  parametrized steady incompressible {N}avier--{S}tokes equations.
\newblock {\em Int. J. Numer. Meth. Engng.}, 102:1136--1161, 2015.

\bibitem{BB01}
R.~Becker and M.~Braack.
\newblock A finite element pressure gradient stabilization for the {S}tokes
  equations based on local projections.
\newblock {\em Calcolo}, 38(4):173--199, 2001.

\bibitem{bergmann2009enablers}
M.~Bergmann, C.~H. Bruneau, and A.~Iollo.
\newblock {Enablers for robust POD models}.
\newblock {\em J. Comput. Phys.}, 228(2):516--538, 2009.

\bibitem{brooks1982streamline}
A.~N. Brooks and T.~J.~R. Hughes.
\newblock Streamline upwind/{P}etrov-{G}alerkin formulations for convection
  dominated flows with particular emphasis on the incompressible
  {N}avier-{S}tokes equations.
\newblock {\em Computer methods in applied mechanics and engineering},
  32(1):199--259, 1982.

\bibitem{caiazzo2014numerical}
A.~Caiazzo, T.~Iliescu, V.~John, and S.~Schyschlowa.
\newblock A numerical investigation of velocity-pressure reduced order models
  for incompressible flows.
\newblock {\em J. Comput. Phys.}, 259:598--616, 2014.

\bibitem{rebollo2015variational}
T.~Ch{\'a}con~Rebollo and B.~M. Dia.
\newblock A variational multi-scale method with spectral approximation of the
  sub-scales$:$ {A}pplication to the {1D} advection--diffusion equations.
\newblock {\em Computer Methods in Applied Mechanics and Engineering},
  285:406--426, 2015.

\bibitem{ChaconCMAME22}
T.~Chac{{\'o}}n~Rebollo, S.~Rubino, M.~Oulghelou, and C.~Allery.
\newblock Error analysis of a residual-based stabilization-motivated {POD-ROM}
  for incompressible flows.
\newblock {\em Comput. Methods Appl. Mech. Engrg.}, 401:115627, 2022.

\bibitem{christie1976finite}
I.~Christie, D.~F. Griffiths, A.~R. Mitchell, and O.~C. Zienkiewicz.
\newblock Finite element methods for second order differential equations with
  significant first derivatives.
\newblock {\em International Journal for Numerical Methods in Engineering},
  10(6):1389--1396, 1976.

\bibitem{CSB03}
M.~Couplet, P.~Sagaut, and C.~Basdevant.
\newblock Intermodal energy transfers in a proper orthogonal
  decomposition--{G}alerkin representation of a turbulent separated flow.
\newblock {\em J. Fluid Mech.}, 491:275--284, 2003.

\bibitem{eroglu2017modular}
F.~G. Eroglu, S.~Kaya, and L.~G. Rebholz.
\newblock A modular regularized variational multiscale proper orthogonal
  decomposition for incompressible flows.
\newblock {\em Comput. Meth. Appl. Mech. Eng.}, 325:350--368, 2017.

\bibitem{giere2015supg}
S.~Giere, T.~Iliescu, V.~John, and D.~Wells.
\newblock {SUPG} reduced order models for convection-dominated
  convection-diffusion-reaction equations.
\newblock {\em Comput. Methods Appl. Mech. Engrg.}, 289:454--474, 2015.

\bibitem{HLB96}
P.~Holmes, J.~L. Lumley, and G.~Berkooz.
\newblock {\em Turbulence, Coherent Structures, Dynamical Systems and
  Symmetry}.
\newblock Cambridge, 1996.

\bibitem{Hughes98}
T.~J.~R. Hughes, G.~R. Feij{{\'o}}o, L.~Mazzei, and J.-B. Quincy.
\newblock The variational multiscale method---a paradigm for computational
  mechanics.
\newblock {\em Comput. Methods Appl. Mech. Engrg.}, 166(1-2):3--24, 1998.

\bibitem{iliescu2013variational}
T.~Iliescu and Z.~Wang.
\newblock Variational multiscale proper orthogonal decomposition:
  {C}onvection-dominated convection-diffusion-reaction equations.
\newblock {\em Math. Comput.}, 82(283):1357--1378, 2013.

\bibitem{iliescu2014variational}
T.~Iliescu and Z.~Wang.
\newblock Variational multiscale proper orthogonal decomposition:
  {N}avier-{S}tokes equations.
\newblock {\em Num. Meth. P.D.E.s}, 30(2):641--663, 2014.

\bibitem{ivagnes2023pressure}
A.~Ivagnes, G.~Stabile, A.~Mola, T.~Iliescu, and G.~Rozza.
\newblock Pressure data-driven variational multiscale reduced order models.
\newblock {\em J. Comput. Phys.}, page 111904, 2023.

\bibitem{john2022error}
V.~John, B.~Moreau, and J.~Novo.
\newblock {Error analysis of a SUPG-stabilized POD-ROM method for
  convection-diffusion-reaction equations}.
\newblock {\em Comput. Math. Appl.}, 122:48--60, 2022.

\bibitem{koc2022verifiability}
B.~Koc, C.~Mou, H.~Liu, Z.~Wang, G.~Rozza, and T.~Iliescu.
\newblock Verifiability of the data-driven variational multiscale reduced order
  model.
\newblock {\em Journal of Scientific Computing}, 93(2):54, 2022.

\bibitem{kragel2005streamline}
B.~Kragel.
\newblock {\em Streamline diffusion {POD} models in optimization}.
\newblock PhD thesis, Universit{\"a}t Trier, 2005.

\bibitem{mou2021data}
C.~Mou, B.~Koc, O.~San, L.~G. Rebholz, and T.~Iliescu.
\newblock Data-driven variational multiscale reduced order models.
\newblock {\em Computer Methods in Applied Mechanics and Engineering},
  373:113470, 2021.

\bibitem{NovoRubinoSINUM21}
J.~Novo and S.~Rubino.
\newblock Error analysis of proper orthogonal decomposition stabilized methods
  for incompressible flows.
\newblock {\em SIAM J. Numer. Anal.}, 59(1):334--369, 2021.

\bibitem{pacciarini2014stabilized}
P.~Pacciarini and G.~Rozza.
\newblock Stabilized reduced basis method for parametrized advection--diffusion
  {PDE}s.
\newblock {\em Comput. Meth. Appl. Mech. Eng.}, 274:1--18, 2014.

\bibitem{parish2017unified}
E.~J. Parish and K.~Duraisamy.
\newblock {A unified framework for multiscale modeling using the Mori-Zwanzig
  formalism and the variational multiscale method}.
\newblock {\em arXiv preprint, \url{http://arxiv.org/abs/1712.09669}}, 2017.

\bibitem{reyes2020projection}
R.~Reyes and R.~Codina.
\newblock {Projection-based reduced order models for flow problems: A
  variational multiscale approach}.
\newblock {\em Comput. Methods Appl. Mech. Engrg.}, 363:112844, 2020.

\bibitem{roop2014proper}
J.~P. Roop.
\newblock A proper-orthogonal decomposition variational multiscale
  approximation method for a generalized oseen problem.
\newblock {\em Advances in Numerical Analysis}, 2014.

\bibitem{RubinoSINUM20}
S.~Rubino.
\newblock Numerical analysis of a projection-based stabilized {POD}-{ROM} for
  incompressible flows.
\newblock {\em SIAM J. Numer. Anal.}, 58(4):2019--2058, 2020.

\bibitem{san2022variational}
O.~San, S.~Pawar, and A.~Rasheed.
\newblock {Variational multiscale reinforcement learning for discovering
  reduced order closure models of nonlinear spatiotemporal transport systems}.
\newblock {\em Sci. Rep.}, 12(1):17947, 2022.

\bibitem{snyder2023data}
W.~Snyder, J.~A. McGuire, C.~Mou, D.~A. Dillard, T.~Iliescu, and R.~De~Vita.
\newblock Data-driven variational multiscale reduced order modeling of vaginal
  tissue.
\newblock {\em Int. J. Num. Meth. Biomed. Eng.}, 39(1):e3660, 2023.

\bibitem{stabile2019reduced}
G.~Stabile, F.~Ballarin, G.~Zuccarino, and G.~Rozza.
\newblock A reduced order variational multiscale approach for turbulent flows.
\newblock {\em Adv. Comput. Math.}, pages 1--20, 2019.

\bibitem{stabile2018finite}
G.~Stabile and G.~Rozza.
\newblock {Finite volume POD-Galerkin stabilised reduced order methods for the
  parametrised incompressible Navier-Stokes equations}.
\newblock {\em Comput. \& Fluids}, 173:273--284, 2018.

\bibitem{tello2019fluid}
A.~Tello, R.~Codina, and J.~Baiges.
\newblock Fluid structure interaction by means of variational multiscale
  reduced order models.
\newblock {\em Int. J. Num. Meth. Eng.}, 2019.

\bibitem{volkwein2013proper}
S.~Volkwein.
\newblock Proper orthogonal decomposition: Theory and reduced-order modelling.
\newblock {\em Lecture Notes, University of Konstanz}, 2013.
\newblock
  \url{http://www.math.uni-konstanz.de/numerik/personen/volkwein/teaching/POD-Book.pdf}.

\bibitem{xie2018data}
X.~Xie, M.~Mohebujjaman, L.~G. Rebholz, and T.~Iliescu.
\newblock Data-driven filtered reduced order modeling of fluid flows.
\newblock {\em SIAM J. Sci. Comput.}, 40(3):B834--B857, 2018.

\bibitem{zoccolan2023streamline}
F.~Zoccolan, M.~Strazzullo, and G.~Rozza.
\newblock {A streamline upwind Petrov-Galerkin reduced order method for
  advection-dominated partial differential equations under optimal control}.
\newblock {\em arXiv preprint, \url{http://arxiv.org/abs/arXiv:2301.01973}},
  2023.

\bibitem{zoccolan2023stabilized}
F.~Zoccolan, M.~Strazzullo, and G.~Rozza.
\newblock Stabilized weighted reduced order methods for parametrized
  advection-dominated optimal control problems governed by partial differential
  equations with random inputs.
\newblock {\em arXiv preprint, \url{http://arxiv.org/abs/arXiv:2301.01975}},
  2023.

\end{thebibliography}

%%%%%%%%%%%%%%%%%%%%%%
\section{Statements and Declarations}
\subsection{Funding}
This work has been supported by Programma Operativo FEDER Andalucia 2014-2020 grant US-1254587 and European Union's Horizon 2020 research, innovation program under the Marie Sklodowska-Curie Actions, grant agreement 872442 (ARIA), Project PID2021-123153OB-C21 funded by MCIN/AEI/10.13039/501100011033/FEDER, UE, and National Science Foundation grant DMS-2012253.

\subsection{Competing Interests}
We wish to confirm that there are no known conflicts of interest associated with this publication and there has been no significant financial support for this work that could have influenced its outcome. Furthermore, the authors have no relevant financial or non-financial interests to disclose.

\subsection{Data Availability}
Data will be made available upon request.

\subsection{Author Contributions}
We confirm that the manuscript has been read and approved by all named authors and that there are no other persons who satisfied the criteria for authorship but are not listed. We further confirm that all have approved the order of authors listed in the manuscript of us. 

We confirm that we have given due consideration to the protection of intellectual property associated with this work and that there are no impediments to publication, including the timing of publication, with respect to intellectual property. In so doing we confirm that we have followed the regulations of our institutions concerning intellectual property.

We understand that the Corresponding Author is the sole contact for the Editorial process (including Editorial Manager and direct communications with the office). The corresponding author is responsible for communicating with the other authors about progress, submissions of revisions, and final approval of proofs. We confirm that we have provided a current, correct email address accessible by the Corresponding Author.

\end{document}